\documentclass[aps,prb,groupedaddress,longbibliography,10pt
]{revtex4-1} 
\usepackage{graphicx,graphics,amsfonts,amssymb,color,bm,wasysym}
\usepackage{comment}

\def\be{\begin{equation}} 
\def\ee{\end{equation}} 
\def\ba{\begin{eqnarray}} 
\def\ea{\end{eqnarray}} 
\def\bc{\begin{center}} 
\def\ec{\end{center}}

\def\p{\partial}

\begin{document} 

\title{Theory of the strongly nonlinear electrodynamic response of graphene: A hot electron model} 

\author{S. A. Mikhailov} 
\email[Electronic mail: ]{sergey.mikhailov@physik.uni-augsburg.de} 

\affiliation{Institute of Physics, University of Augsburg, D-86135 Augsburg, Germany} 

\date{\today} 

\begin{abstract}
An electrodynamic response of graphene to a strong electromagnetic radiation is considered. A hot electron model (HEM) is introduced and a corresponding system of nonlinear equations is formulated. Solutions of this system are found and discussed in detail for intrinsic and doped graphene: the hot electron temperature, non-equilibrium electron and holes densities, absorption coefficient and other physical quantities are calculated as functions of the incident wave frequency $\omega$ and intensity $I$, of the equilibrium chemical potential $\mu_0$ and temperature $T_0$, scattering parameters, as well as of the ratio $\tau_\epsilon/\tau_{\rm rec}$ of the intra-band energy relaxation time $\tau_\epsilon$ to the recombination time $\tau_{\rm rec}$. The influence of the radiation intensity on the absorption coefficient $A$ at low ($\hbar\omega\lesssim 2|\mu_0|$, $dA/dI>0$) and high ($\hbar\omega\gtrsim 2|\mu_0|$, $dA/dI<0$) frequencies is studied. The results are shown to be in good agreement with recent experimental data. 
\end{abstract} 


 
\maketitle 

\tableofcontents

\section{Introduction\label{sec:intro}}

The nonlinear electrodynamic response of graphene attracted great attention in recent years. After the pioneering prediction \cite{Mikhailov07e} of the strongly nonlinear electrodynamic properties of graphene, a large number of theoretical \cite{Mikhailov08a,Dean09,Dean10,Smirnova14,Savostianova15,Wang16,Rostami16a, Cheng17,Savostianova17a,Mikhailov12b,Yao14,Tokman16,Cheng14b,MariniAbajo17,Savostianova18a,Mikhailov09b, Yao13,Tokman14,Mikhailov11c,Mikhailov12c,Peres14,CoxAbajo14,CoxAbajo15,CoxAbajo16,Mikhailov17b,Mikhailov17a, Cheng14a,Cheng15,Mikhailov16a,Mikhailov09a,Ishikawa10,Cheng15b,Semnani16,Mikhailov19} and experimental 
\cite{Dragoman10,Bykov12,Kumar13,Hong13,Soavi18,Hendry10,Gu12,Alexander17, KonigOtto17,Alexander18,Bao09,Zhang09,Winnerl11,Zheng12,Bianchi17,Zhang12,Chen13,Miao15, Dremetsika16,Vermeulen16,Tomadin18} papers have been published. 
Theoretically the higher harmonics generation \cite{Mikhailov07e,Mikhailov08a,Dean09,Dean10,Smirnova14,Savostianova15,Wang16,Rostami16a, Cheng17,Savostianova17a}, the frequency mixing \cite{Mikhailov12b,Yao14,Tokman16}, the direct current induced second harmonic generation \cite{Cheng14b}, the saturable absorption and Kerr effects \cite{MariniAbajo17,Savostianova18a} have been studied. The nonlinear graphene response in magnetic fields \cite{Mikhailov09b,Yao13,Tokman14} and the plasma wave related nonlinear effects \cite{Mikhailov11c,Mikhailov12c,Yao14,CoxAbajo14,CoxAbajo15, CoxAbajo16,Mikhailov17b} have been also discussed in detail. A nonperturbative quasiclassical theory based on the relaxation time approximation and a quantum perturbation theory of all third order nonlinear effects have been developed in Ref. \cite{Mikhailov17a} and in Refs. \cite{Cheng14a,Cheng15,Mikhailov16a} respectively. Experimentally the higher harmonics generation \cite{Dragoman10,Bykov12,Kumar13,Hong13,Soavi18}, the four-wave mixing \cite{Hendry10,Gu12,Alexander17,KonigOtto17,Alexander18}, the radiation induced  absorption changes\cite{Bao09,Zhang09,Winnerl11,Zheng12,Bianchi17}, Kerr effect \cite{Zhang12,Chen13,Miao15,Dremetsika16,Vermeulen16}, the photoconductivity \cite{Tomadin18} and other nonlinear phenomena have been observed. All of them demonstrated very large absolute values of the nonlinear optical parameters of graphene. The nonlinear electrodynamic properties of graphene can be used in many applications including broadband detection \cite{Ryzhii12,Gan13,Wang13,Pospischil13,Ryzhii15}, electrically tunable modulation of terahertz \cite{Liu15,Kindness18} and optical radiation \cite{Liu11,Phare15}, mode-locked lasers \cite{Sun10,Zhang10,Popa10,Popa11,Bao11} and other \cite{Ryzhii07,Otsuji12,Ryzhii13}.

The third-order fourth-rank conductivity tensor of graphene $\sigma_{\alpha\beta\gamma\delta}^{(3)}(\omega_1,\omega_2,\omega_3;\mu_0,T_0)$, analytically calculated in the quantum theory \cite{Cheng14a,Cheng15,Mikhailov16a}, describes all possible third-order nonlinear effects for arbitrary polarizations and frequencies of the incident waves. Since $\sigma_{\alpha\beta\gamma\delta}^{(3)}$ was calculated within the perturbation theory, it depends, apart from the input frequencies $\omega_1$, $\omega_2$, $\omega_3$, on the equilibrium chemical potential $\mu_0$ and equilibrium temperature $T_0$. In many nonlinear response experiments, however, the incident radiation is so strong that the system gets excited far beyond the equilibrium state, and the use of parameters $\mu_0$, $T_0$ becomes not fully relevant. This problem can be partly circumvented by replacing $T_0$ in the expression for $\sigma_{\alpha\beta\gamma\delta}^{(3)}$ by an effective temperature $T$ which is considered as a fitting parameter and can be (much) larger than $T_0$; this way to interpret experimental data was used, e.g., in Refs. \cite{Winnerl11,Soavi18,Alexander18,Tomadin18}. However, in general, not only the temperature, but also the chemical potentials of electron ($\mu_e$) and hole ($\mu_h$) gases should be considered to be different from $\mu_0$. 

The description of a strongly nonequilibrium electron-hole plasma in terms of the quasi-equilibrium electron and hole Fermi gases with their own chemical potentials and temperatures \cite{Malic11,Sun12,Song13,Tomadin13} is justified if the electron-electron, electron-hole and hole-hole scattering processes (characterized by a typical scattering time $\tau_{ee}$) are more probable than the electron-phonon and electron-impurities ones. There exist theoretical arguments \cite{Song13,Tomadin13} and experimental evidences \cite{Lui10,Breusing11,Brida13,Johannsen13} that in typical graphene samples this situation is the case indeed. 

Although the hot electron model (HEM) has been already used for interpretation of several nonlinear graphene experiments, a comprehensive theory which would analyze different physical situations and would give the opportunity to calculate $\mu_e$, $\mu_h$ and $T$ as a function of different input parameters of the problem is still absent. In this paper we develop such a theory. In Section \ref{sec:theor} we introduce a HEM and formulate a system of nonlinear differential equations which allows to calculate $\mu_e$,  $\mu_h$, $T$ and other physical quantities characterizing the electron-hole plasma in graphene in the strongly non-equilibrium state. In Section \ref{sec:results} we analyze solutions of this system of equations in doped and intrinsic graphene, as well as compare results of our theory with some experimental data. In Section \ref{sec:conclusion} the results are summarized and conclusions are drawn.

\section{Theory\label{sec:theor}}

\subsection{The system in equilibrium}

We consider a graphene monolayer lying at the plane $z=0$ on top of a dielectric with the dielectric constant $\kappa$ and the refractive index $n=\sqrt{\kappa}$. The energy spectrum of electrons ($l=2$) and holes ($l=1$) in graphene is
\be 
E_{l{\bm k}}=(-1)^l\hbar v_F |\bm k|,
\ee
where $v_F\approx 10^8$ cm/s is the Fermi velocity and the energy $E_{l{\bm k}}$ and the wave vector $\bm k$ are counted from one of the Dirac points. In equilibrium (without irradiation) the electron distribution function has the form (the Boltzmann constant $k_B=1$ is set to be unity everywhere)
\be 
f_0(E,\mu_0,T_0)=\frac 1{1+\exp\left(\frac{E-\mu_0}{T_0}\right)},
\label{Distrfunc0}
\ee
where $\mu_0$ and $T_0$ are the equilibrium chemical potential and temperature, the same for electrons and holes. 

Below we will analyze two representative cases, with $\mu_0=-0.2$ eV (doped graphene) and $\mu_0=0$ eV (intrinsic graphene). If $\mu_0=-0.2$ eV then at room temperature $T_0=300$ K the equilibrium densities of electrons and holes,
\be 
n_e^0=4.35\times 10^7\textrm{ cm}^{-2},\ \ \  
n_h^0=3.12\times 10^{12}\textrm{ cm}^{-2},\ \ \textrm{(doped)}
\label{dens-example}
\ee
differ by almost five orders of magnitude. In intrinsic graphene at $T_0=300$ K the densities are
\be 
n_e^0=n_h^0=8.14\times 10^{10}\textrm{ cm}^{-2},\ \ \textrm{(intrinsic)}.
\label{dens-example-2}
\ee
The equilibrium chemical potential  $\mu_0$ can be experimentally varied by the gate voltage.

\subsection{Hot electrons distribution functions}

Now we assume that graphene is irradiated by an external electromagnetic wave with the frequency $\omega$ and intensity $I$. The photon energy $\hbar\omega$ can be both larger and smaller than $2|\mu_0|$, and the intensity of radiation is assumed to be so large that the perturbation theory is inapplicable. The photo-excited electrons absorb the wave energy, due to the intra- and inter-band absorption processes, and relax their energy to the crystal lattice and to the substrate via different scattering processes. We denote the electron-electron (as well as hole-hole and electron-hole) scattering time as $\tau_{ee}$, the momentum and energy \textit{intra}-band relaxation times, due to the electron scattering by lattice imperfections (phonons, impurities, etc.), as $\tau_p$ and $\tau_\epsilon$, and the \textit{inter}-band energy relaxation (actually recombination) time as $\tau_{\rm rec}$ (the time $\tau_{\rm rec}$ will be discussed later in Section \ref{DynEq}). The momentum relaxation time $\tau_p$ is typically much smaller than $\tau_\epsilon$, $\tau_p\ll\tau_\epsilon$. Further, we will accept a hypothesis \cite{Song13,Tomadin13,Lui10,Breusing11,Brida13,Johannsen13} that the electron-electron scattering time is smaller than $\tau_p$, 
\be 
\tau_{ee}\ll\tau_p\ll\tau_\epsilon ;
\label{comparetimes}
\ee
according to the literature, $\tau_{ee}$ is about a few tens of fs, while $\tau_p$ is at least 0.1 ps or larger. Under these conditions, shortly after the excitation \textit{quasi}-equilibrium Fermi distributions 
\be 
f(E,\mu_e,\mu_h,T)=
\frac {\Theta(E)}{1+\exp\left(\frac{E-\mu_e}T\right)}+
\frac {\Theta(-E)}{1+\exp\left(\frac{E-\mu_h}T\right)},
\label{f_glob}
\ee
with the electron ($\mu_e$) and hole ($\mu_h$) chemical potentials and the common temperature $T\neq T_0$, are formed in the conduction and valence bands. The distribution functions of electron ($f_e=f$) and holes ($f_h=1-f$) then read
\be 
f_e(E,\mu_e,T)=\frac 1{1+\exp\left(\frac{E-\mu_e}{T}\right)}, \ E>0,
\label{f_e}
\ee
\be 
f_h(E,\mu_h,T)=\frac 1{1+\exp\left(\frac{\mu_h-E}{T}\right)}, \ E<0.
\label{f_h}
\ee
It is also possible to consider the version of the theory in which the temperatures of the electron and hole gases, $T_e$ and $T_h$, are different. This corresponds to a situation in which electron-electron and hole-hole scattering is more likely than electron-hole scattering. As was shown in Ref. \cite{Sun12} this is typically not the case, therefore we will restrict ourselves by the model with $T_e=T_h=T$.

We have introduced three unknown quantities $\mu_e$, $\mu_h$ and $T$, and now need equations which would determine their dependencies on the equilibrium parameters $\mu_0$ and $T_0$, as well as on the frequency and intensity of the incident radiation.

\subsection{Electron and hole densities}

The density of electrons and holes in the strongly non-equilibrium state (\ref{f_glob}) are determined by the distribution functions (\ref{f_e})--(\ref{f_h}) in the usual way, 
\be 
n_e(\mu_e,T)=\frac {g_sg_v}{S}\sum_{\bm k}f_e(E_{2{\bm k}},\mu_e,T)=
\frac {2T^2}{\pi(\hbar v_F)^2}F_1\left(\frac{\mu_e}T\right) ,
\label{densN0}
\ee
\be 
n_h(\mu_h,T)=
\frac {2T^2}{\pi(\hbar v_F)^2}F_1\left(-\frac{\mu_h}T\right),\label{densP0}
\ee
where $g_s=g_v=2$ are the spin and valley degeneracies, $S$ is the sample area, and the function $F_n(z)$ is defined as
\be 
F_n(z)=\int_0^\infty  \frac {x^n dx}{1+\exp\left(x-z\right)} .\label{funF}
\ee
The equilibrium electron and hole densities (\ref{dens-example})--(\ref{dens-example-2}) are determined by Eqs. (\ref{densN0})--(\ref{densP0}) in which $\mu_e=\mu_h=\mu_0$ and $T=T_0$.

\subsection{Electron and hole energy densities}

The energy density of the electron and hole gases per unit area (per cm$^2$) is determined by
\be 
{\cal E}_e (\mu_e,T)=\frac {g_sg_v}{S}\sum_{\bm k}E_{2{\bm k}}f_e(E_{2{\bm k}},\mu_e,T)
=\frac {2T^3}{\pi(\hbar v_F)^2}F_2\left(\frac{\mu_e}T\right)
\label{energyE}
\ee
\be 
{\cal E}_h (\mu_h,T)
=\frac {2T^3}{\pi(\hbar v_F)^2}F_2\left(-\frac{\mu_h}T\right).
\label{energyH}
\ee
The total energy of electrons and holes is
\ba 
{\cal E} (\mu_e,\mu_h,T)={\cal E}_e (\mu_e,T)+{\cal E}_h (\mu_h,T)
\ea

\subsection{Conductivity}

The linear-response conductivity has three contributions (the derivation can be found, e.g., in Ref. \cite{Mikhailov16a}): intra-band electron, intra-band hole and inter-band,
\be 
\sigma^{(1)}(\omega,\mu_e,\mu_h,T)=
\sigma^{(1),e}_{\rm intra}(\omega,\mu_e,T)+
\sigma^{(1),h}_{\rm intra}(\omega,\mu_h,T)+
\sigma^{(1)}_{\rm inter}(\omega,\mu_e,\mu_h,T),
\label{cond-contribs}
\ee
where
\be
\frac{\sigma^{(1),e}_{\rm intra}(\omega,\mu_e,T)}{\frac{e^2g_sg_v}{16\hbar}}
=
\frac {i}{\pi \hbar T}\int_0^\infty   
\frac {EdE}{\omega+i\gamma_{\rm intra}(E)}
\frac 1{\cosh^2\left(\frac{E-\mu_e}{2T}\right)},
\label{cond-intra-e}
\ee
\be
\frac{\sigma^{(1),h}_{\rm intra}(\omega,\mu_h,T)}{\frac{e^2g_sg_v}{16\hbar}}
=
\frac {i}{\pi \hbar T}\int_0^\infty   
\frac {EdE}{\omega+i\gamma_{\rm intra}(E)}
\frac 1{\cosh^2\left(\frac{E+\mu_h}{2T}\right)},
\label{cond-intra-h}
\ee
and
\ba
\frac{\sigma^{(1)}_{\rm inter}(\omega,\mu_e,\mu_h,T)}{\frac {e^2g_sg_v}{16\hbar}}
=
\frac {-i}{\pi}
\int_{0}^\infty dE
\left(\frac 1{1+\exp\left(\frac{-E-\mu_h}T\right)}-\frac 1{1+\exp\left(\frac{E-\mu_e}T\right)}\right)
\frac{\hbar(\omega+i\gamma_{\rm inter})} { E^2-[\hbar(\omega+i\gamma_{\rm inter})/2]^2} .\label{cond-inter}
\ea
We discuss these contributions separately.

\subsubsection{Intra-band conductivity}

In order to calculate the intra-band conductivity (\ref{cond-intra-e})--(\ref{cond-intra-h}) we need a model for the scattering rate $\gamma_{\rm intra}(E)\equiv 1/\tau_p(E)$, where $\tau_p(E)$, see Eq. (\ref{comparetimes}), is the energy-dependent momentum relaxation time due to the scattering of electrons and holes with impurities, phonons and other lattice imperfections (but not with each other). As a first choice we use for $\gamma_{\rm intra}(E)$ the model
\be 
\hbar\gamma_{\rm intra}(E)=\frac{|E|}{\frac \zeta 2+\sqrt{1+\frac{E^4}{ E_{i}^4}}-1} \label{gam-all-E}
\ee
which is discussed in detail in Appendix \ref{app:gamma-intra}. The quantities $\zeta$ and $E_i$ in (\ref{gam-all-E}) are fitting parameters: $\zeta$ is the minimal static conductivity of graphene in the Dirac point, in units $e^2/h$; $E_i\propto \sqrt{N_i}$ is a Coulomb energy associated with the density of impurities $N_i$, see (\ref{coul-en-1}). The parameters $\zeta$ and $E_i$ can be found by fitting the formula (\ref{sigma-fit}) to the experimental data for the gate voltage dependence of the \textit{static} linear conductivity of graphene, see example in Figure \ref{fig:Chen08}. Thus found parameters $\zeta$ and $E_i$ and the model expression (\ref{gam-all-E}) are then used in formulas (\ref{cond-intra-e})--(\ref{cond-intra-h}) for the \textit{high-frequency} nonlinear conductivities of graphene. In the rest of the paper we use $\zeta=4$ and $E_i=30$ meV, which corresponds to the mobility of about 7260 cm$^2$/Vs. For the relation between the energy $E_i$ and the sample mobility, as well as for further discussion of the model (\ref{gam-all-E}) see Appendix \ref{app:gamma-intra}. 

Alternatively, we also use the energy-independent momentum relaxation rate model with $\gamma_{\rm intra}=\gamma_{p}=1/\tau_p$. Then the intra-band dynamic conductivity assumes the form
\be
\frac{\sigma^{(1)}_{intra}(\omega,\mu_e,\mu_h,T)}{\frac{e^2g_sg_v}{16\hbar}}
=\frac {4i}{\pi }\frac T{\hbar (\omega+i \gamma_{p})}
\left[F_0\left(\frac{\mu_e}{T}\right)
+F_0\left(-\frac{\mu_h}{T}\right)
\right]=\frac {4i}{\pi }\frac T{\hbar (\omega+i \gamma_{p})}
\ln\left[\left(1+e^{\mu_e/T}\right)\left(1+e^{-\mu_h/T}\right)
\right].\label{GammaEindep}
\ee

The model (\ref{gam-all-E}) better reproduces typical experimental data on the static conductivity of graphene, therefore we use it in the main part of the paper. The model (\ref{GammaEindep}) was used in some experiments (e.g., Ref. \cite{Winnerl11}) by interpreting the measured data; comparing our results with Ref. \cite{Winnerl11} in Section \ref{sec:results} we also use the model (\ref{GammaEindep}) with the energy-independent momentum relaxation rate $\gamma_{\rm intra}$.

\subsubsection{Inter-band dynamic conductivity}

The inter-band conductivity (\ref{cond-inter}) depends on the inter-band scattering rate $\gamma_{\rm inter}$. We assume that $\gamma_{\rm inter}\to 0$ since a finite $\gamma_{\rm inter}$ does not influence the final result under the condition $\hbar\gamma_{\rm inter}\ll T$ which is typically satisfied in experiments. Then Eq. (\ref{cond-inter}) can be simplified so that the real part assumes the form 
\be
\textrm{Re }\frac{\sigma^{(1)}_{\rm inter}(\omega,\mu_e,\mu_h,T)}{\frac {e^2g_sg_v}{16\hbar}}
=
\frac {
\sinh\left(\frac{\hbar|\omega|-(\mu_e-\mu_h)}{2T}\right)}
{\cosh\left(\frac{\mu_h+\mu_e}{2T}\right)+ \cosh\left(\frac{\hbar|\omega|-(\mu_e-\mu_h)}{2T}\right) }
\ee
and the imaginary part is expressed in terms of a principal value integral (denoted by ${\cal P}$),
\be
\textrm{Im }\frac{\sigma^{(1)}_{\rm inter}(\omega,\mu_e,\mu_h,T)}{\frac {e^2g_sg_v}{16\hbar}}
=
-\frac {\hbar\omega}{\pi}{\cal P}
\int_{0}^\infty 
\frac {
\sinh\left(\frac{E-(\mu_e-\mu_h)/2}T\right)}
{\cosh\left(\frac{\mu_h+\mu_e}{2T}\right)+ \cosh\left(\frac{E-(\mu_e-\mu_h)/2}T\right) }
\frac{dE}
{ E^2-(\hbar\omega/2)^2} .
\ee
Notice that the real part of the inter-band conductivity can be negative if $\mu_e>\mu_h$ and $\hbar\omega<(\mu_e-\mu_h)$. Physically this is due to the population inversion in the non-equilibrium state.

\subsection{Absorption coefficient}

We assume that graphene lies on the surface of a dielectric substrate with the dielectric constant $\kappa$ and the refractive index $n=\sqrt{\kappa}$, and the external radiation with the intensity $I$ is normally incident on the structure. The incident radiation is transmitted through (with the intensity $TI$), reflected from (the intensity $RI$) and absorbed in the graphene layer (the intensity $AI$). The absorbed part of the radiation energy is determined by the absorption coefficient $A$, sometimes also referred to as absorbance. The coefficient $A$ is determined by the Joule heating $\bm j\cdot \bm E$ and in the linear-response regime is proportional to the real part of the first-order conductivity $\sigma^{(1)}$. In the nonlinear regime we will assume that the nonlinearity mainly manifests itself in changing the chemical potentials ($\mu_0\to\mu_e,\mu_h$) and electron temperature ($T_0\to T$) in formulas (\ref{cond-contribs}) -- (\ref{cond-inter}) and hence, in accordance with these equations, the absorption coefficient can be presented in the form 
\be 
A(\omega,\mu_e,\mu_h,T)=A^{\rm intra}(\omega,\mu_e,\mu_h,T)+ A^{\rm inter}(\omega,\mu_e,\mu_h,T),
\label{absorptn-contr}
\ee
with intra-,
\be 
A^{\rm intra}(\omega,\mu_e,\mu_h,T)=A^{\rm intra}_e(\omega,\mu_e,T)+A^{\rm intra}_h(\omega,\mu_h,T)=\frac{\frac{4\pi}c\textrm{Re }\sigma^{(1)}_{\rm intra}(\omega,\mu_e,\mu_h,T)}
{\left|\frac{n+1}2+\frac{2\pi}c\sigma^{(1)}(\omega,\mu_e,\mu_h,T)\right|^2},
\label{absorptn-intra}
\ee 
and inter-band, 
\be 
A^{\rm inter}(\omega,\mu_e,\mu_h,T)=\frac{\frac{4\pi}c\textrm{Re }\sigma^{(1)}_{\rm inter}(\omega,\mu_e,\mu_h,T)}
{\left|\frac{n+1}2+\frac{2\pi}c\sigma^{(1)}(\omega,\mu_e,\mu_h,T)\right|^2},
\label{absorptn-inter}
\ee
contributions. The denominators in Eqs. (\ref{absorptn-intra})--(\ref{absorptn-inter}) contain the total conductivity. If the substrate is made out of silicon dioxide then its dielectric constant is $\kappa_{\rm{SiO}_2} = 3.9$ and $n=\sqrt{\kappa_{\rm{SiO}_2}}=1.975$.  

The absorption coefficient formulas (\ref{absorptn-contr}) -- (\ref{absorptn-inter}) are approximate. In general the current $\bm j$ contains the higher contributions $j_\alpha^{(3)}=\sigma^{(3)}_{\alpha\beta\gamma\delta}E_\beta E_\gamma E_\delta$, $j_\alpha^{(5)}=\sigma^{(5)}_{\alpha\beta\gamma\delta\mu\nu}E_\beta E_\gamma E_\delta E_\mu E_\nu$, etc., where all higher-order conductivities should be considered as functions of non-equilibrium chemical potentials $\mu_e$, $\mu_h$ and temperature $T$. However at present the functions $\sigma^{(5)}$, $\sigma^{(7)}$, etc., are unknown and the function $\sigma^{(3)}_{\alpha\beta\gamma\delta}(\omega_1,\omega_2,\omega_3;\mu_0,T_0)$ was calculated \cite{Cheng14a,Cheng15,Mikhailov16a} only for the quasi-equilibrium case with $\mu_0$ and $T_0$. Therefore in this paper we restrict ourselves by the approach (\ref{absorptn-contr}) -- (\ref{absorptn-inter}), 
postponing developing of more general theories for future publications.

\subsection{Dynamics equations of the hot electron model \label{DynEq}}

Now we are prepared to formulate the basic equations of our HEM. We will assume that the intra-band energy relaxation time $\tau_\epsilon$ is shorter than the inter-band recombination time $\tau_{\rm rec}$, $\tau_\epsilon\ll\tau_{\rm rec}$. This condition is typically satisfied in conventional semiconductors. In graphene the radiative recombination time, according to estimates in Refs. \cite{Vasko08,Alymov18}, is around hundreds of nanoseconds at room temperature, while $\tau_\epsilon$ lies in the tens-of-ps range. This justifies the use of the condition $\tau_\epsilon\ll\tau_{\rm rec}$ below.

\subsubsection{Energy relaxation}

Assume that the system is excited by a powerful incident radiation with the intensity $I$. Since the recombination is a slow process, the electron and hole Fermi gases are independent from each other in that sense that they are characterized (at the time $t\gtrsim \tau_{ee}$ after the excitation is switched on) by their own chemical potentials $\mu_e$ and $\mu_h$, and the temperature $T$. At the longer time scale $t\sim \tau_\epsilon$ the charge carriers, having been scattered by phonons, impurities and other lattice imperfections, relax their energy to the lattice. We assume that the energy relaxation equations for electrons and holes can then be written, as it is usually done in semiconductor physics, in the form
\be 
\frac{\p {\cal E}_e(\mu_e,T)}{\p t}=A^{\rm intra}_e(\omega,\mu_e,T)I-\frac{{\cal E}_e(\mu_e,T)-{\cal E}_e(\mu_e^0,T_0)}{\tau_\epsilon},
\label{en-relax-e}
\ee
\be 
\frac{\p {\cal E}_h(\mu_h,T)}{\p t}=A^{\rm intra}_h(\omega,\mu_h,T)I-\frac{{\cal E}_h(\mu_h,T)-{\cal E}_h(\mu_h^0,T_0)}{\tau_\epsilon},
\label{en-relax-h}
\ee
meaning that the energy of hot electron and hole gases grows in time due to the intra-band absorption in each (conduction and valence) band and relaxes to their steady-state quasi-equilibrium energies ${\cal E}_e(\mu_e^0,T_0)$ and ${\cal E}_h(\mu_h^0,T_0)$ with the characteristic time scale $\tau_\epsilon$. We emphasize that the temperature of the relaxed quasi-equilibrium state in Eqs. (\ref{en-relax-e})--(\ref{en-relax-h}) coincides with the lattice temperature $T_0$ since $\tau_\epsilon$ describes the relaxation processes between the charge carrier gases and the lattice. However, the chemical potentials $\mu_e^0$ and $\mu_h^0$ differ from the equilibrium chemical potential $\mu_0$ since the density of electrons and holes are still larger than those in equilibrium since $\tau_\epsilon\ll\tau_{\rm rec}$. The relation between $\mu_{e,h}$ and $T$, from one side, and $\mu_{e,h}^0$ and $T_0$, from the other side, is determined by the conservation of the electron and hole densities,
\be 
n_e(\mu_e,T)=n_e(\mu_e^0,T_0), \ \ \ 
n_h(\mu_h,T)=n_h(\mu_h^0,T_0).\label{relation-densities}
\ee
The energy relaxation times $\tau_{\epsilon}$ in Eqs. (\ref{en-relax-e})--(\ref{en-relax-h}) can, in principle, be different. We will assume, for simplicity, that they are the same. Then we can take a sum of Eqs. (\ref{en-relax-e})--(\ref{en-relax-h}) and get the total energy relaxation equation
\be 
\frac{\p {\cal E}(\mu_e,\mu_h,T)}{\p t}=A^{\rm intra}(\omega,\mu_e,\mu_h,T)I-\frac{{\cal E}(\mu_e,\mu_h,T)-{\cal E}(\mu_e^0,\mu_h^0,T_0)}{\tau_\epsilon}.
\label{en-relax}
\ee

\subsubsection{Recombination \label{sec:recomb}}

At a longer time scale $\sim \tau_{\rm rec}$ electrons and holes recombine. Taking into account that they are generated and recombine by pairs, 
\be 
n_e(t)=n_e^0+\delta n(t),
\ \ \  
n_h(t)=n_h^0+\delta n(t),
\label{deltaNdeltaP}
\ee  
we write the generation-recombination rate equation in the form
\be 
\frac{\p n_e}{\p t}=\frac{\p n_h}{\p t}=G-R=\frac{A^{\rm inter}(\omega,\mu_e,\mu_h,T)}{\hbar\omega} I -\alpha_{\rm rec} (n_en_h-n_e^0n_h^0).
\label{GRequation}
\ee
The first (generation) term $G$ in the right hand side of (\ref{GRequation}) represents the number of electron-hole pairs generated per second on a unit area. It equals the radiation intensity $I$ (the radiation energy \textit{incident} on a unit area per second), times the \textit{inter-band} absorption coefficient (which gives the energy \textit{absorbed} on a unit area per second due to the electron-hole generation processes), and divided by the photon energy (which results in the number of electron-hole pairs generated on a unit area per  second). The second term $R$ in the right hand side of (\ref{GRequation}) is the recombination rate. The recombination is a nonlinear bi-particle process with the recombination rate being proportional to the product of electron and hole densities $n_en_h$. The recombination term describes the relaxation to the equilibrium electron and hole densities $n_e^0n_h^0$. The recombination coefficient $\alpha_{\rm rec}$ is measured in units cm$^2$/s and is independent of the particle densities. It is the second (in addition to $\tau_\epsilon$) parameter of the theory.

\begin{figure}[t]
\includegraphics[width=0.49\textwidth]{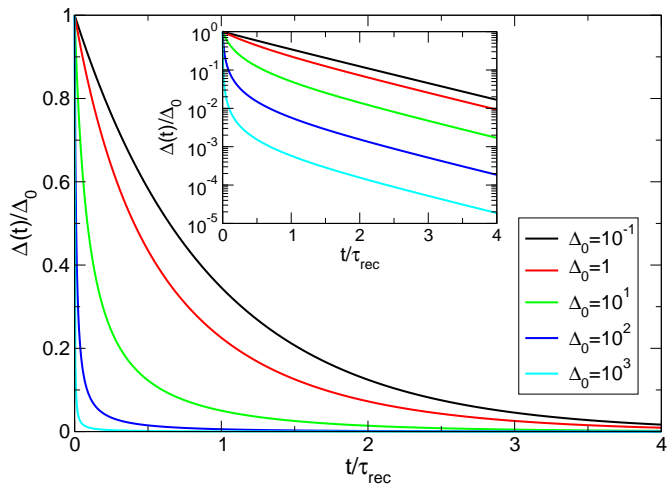}
\caption{\label{fig:recomb}The time dependence  of the normalized electron-hole pairs density $\Delta(t)$, Eq. (\ref{recomb}), at different values of $\Delta_0$. The main plot and the inset show the same curves with linear and logarithmic scales of the $y$-axis respectively. The decay of the electron-hole pairs density is exponential only at low excitation levels $\Delta_0\ll 1$.}
\end{figure}

Apart from the recombination coefficient $\alpha_{\rm rec}$ one can also introduce a quantity $\tau_{\rm rec}$ which is measured in units of time and \textit{at low excitation levels} has the meaning of the recombination time (in general the recombination process is not purely exponential and the meaning of $\tau_{\rm rec}$ is more complicated, see below). Assume that the radiation intensity $I(t)$ is switched off at the time moment $t=0$ and consider the time evolution of the electron and hole densities $\delta n(t)$ at $t>0$. Substituting (\ref{deltaNdeltaP}) into equation (\ref{GRequation}) with $I=0$ and taking into account the initial condition $\delta n(0)=\delta n_0$ we get
\be 
\Delta(t) \equiv \frac{\delta n(t)}{n_e^0+n_h^0}
=\frac{\Delta_0}
{\left(1+\Delta_0\right)e^{ \alpha_{\rm rec}(n_e^0+n_h^0)t }-\Delta_0}
=\frac{\Delta_0}
{\left(1+\Delta_0\right)e^{t/\tau_{\rm rec} }-\Delta_0}
\label{recomb}
\ee
where $\Delta(t)$ is the relative change of the charge carrier density as compared to their total equilibrium density, and $\Delta_0=\Delta(0)$. The quantity 
\be 
\tau_{\rm rec}=\frac 1{\alpha_{\rm rec}(n_e^0+n_h^0)}
\label{taurec}
\ee
has the dimension of time, depends on the total equilibrium density of electrons and holes and determines the time evolution of the electron-hole recombination (\ref{recomb}). We emphasize that $\tau_{\rm rec}$ has the meaning of time over which the initial carrier density reduces by a factor of $\sim 2.72...$ \textit{only} at very low excitation levels $\Delta_0\ll 1$, 
\be 
\Delta (t)\approx \Delta_0e^{- t/\tau_{\rm rec} }, \ \ \Delta_0\ll 1,
\label{recomb-low-excite}
\ee
see Figure \ref{fig:recomb}. At high excitation levels, $\Delta_0\gg 1$ or $\delta n_0\gg (n_e^0+n_h^0)$, the density $\delta n(t)$ first very quickly decreases, with the time constant $\sim \tau_{\rm rec}/\Delta_0$, down to the values $\sim (n_e^0+n_h^0)$, and then decays further exponentially, see inset to Figure \ref{fig:recomb}. Quantitatively, $\Delta (t)$ decreases from its initial value $\Delta_0$ by a factor of two during the time $\tau_{\rm rec}/(\Delta_0+1)\ll \tau_{\rm rec}$ and by a factor of $\Delta_0$ during the time $\sim\tau_{\rm rec}$. 

The generation-recombination rate equation (\ref{GRequation}) can be also rewritten in the form explicitly containing $\tau_{\rm rec}$,
\be 
\frac{\p (\delta n)}{\p t}=\frac{A^{\rm inter}(\omega,\mu_e,\mu_h,T)}{\hbar\omega} I -
\frac {\delta n}{\tau_{\rm rec}}
\left(1+\frac{\delta n}{n_e^0+n_h^0}\right).
\label{recombination}
\ee

\subsubsection{Preliminary summary and discussion}

The recombination of charge carriers in graphene characterized by a more complicated than $e^{-t/\tau}$-decay was commonly observed in time-resolved pump-probe experiments, see, e.g., Refs. \cite{Dawlaty08a,George08,Winnerl11} and other. It was often interpreted by introducing two different time scales $\tau_1$ and $\tau_2$ where different $\tau$-s were attributed to physically different relaxation mechanisms. As seen from Section \ref{sec:recomb} the seemingly double-$\tau$ time decay is actually described by a single formula (\ref{recomb}) with only one decay-time parameter $\tau_{\rm rec}$. The reason of the more complicated behavior of $\delta n(t)$ is the intrinsically nonlinear nature of the electron-hole recombination process seen in Eqs. (\ref{GRequation}), (\ref{recombination}). The corresponding two time constants are $\tau_1= \tau_{\rm rec}/(\Delta_0+1)$ and $\tau_2=\tau_{\rm rec}$. In the strong excitation limit $\Delta_0\gg 1$ the first time is much shorter than the second one, $\tau_1\ll\tau_2$; while in the weak excitation limit ($\Delta_0\ll 1$) the two times merge into one, $\tau_1\simeq\tau_2=\tau_{\rm rec}$, Figure \ref{fig:recomb}.

The nonlinearity of the recombination process is known in the semiconductor physics, e.g., Ref. \cite{Bonch77}. We have briefly reproduced here the nonlinear recombination equations (\ref{GRequation}), (\ref{recombination}) and the corresponding derivation of Eq. (\ref{recomb}) since in some recent papers (e.g., Ref. \cite{Soavi18}) a strongly non-equilibrium ($\Delta_0\gg 1$) recombination dynamics has been improperly described by a linear recombination term $R\propto -\delta n/\tau_{\rm rec}$.

Equations (\ref{en-relax}) and (\ref{recombination}) [or (\ref{en-relax}) and (\ref{GRequation})] describe the dynamics of the electron temperature $T$, chemical potentials $\mu_e$, $\mu_h$, and all other physical quantities within our HEM. The energy densities ${\cal E}_e$, ${\cal E}_h$ in Eq. (\ref{en-relax}), as well as the intra- and inter-band absorption coefficients $A^{\rm intra}$, $A^{\rm inter}$ in Eqs. (\ref{en-relax}) and (\ref{recombination}), depend on six unknown quantities $T$, $\mu_e$, $\mu_h$, $\mu_e^0$, $\mu_h^0$, and $\delta n$. The four missing equations can be found by inverting the relations (\ref{densN0}) and (\ref{densP0}), namely, 
\be 
\mu_e=TG \left(\frac{\pi(\hbar v_F)^2 n_e}{2T^2}\right)
=TG \left(\frac{\pi(\hbar v_F)^2 (n_e^0+\delta n)}{2T^2}\right)
\label{muE}
\ee
\be 
\mu_h=-TG \left(\frac{\pi(\hbar v_F)^2 n_h}{2T^2}\right)
=-TG \left(\frac{\pi(\hbar v_F)^2 (n_h^0+\delta n)}{2T^2}\right)
\label{muH}
\ee
\be 
\mu_e^0=T_0G \left(\frac{\pi(\hbar v_F)^2 (n_e^0+\delta n)}{2T_0^2}\right)
\label{muE0}
\ee
\be 
\mu_h^0=-T_0G \left(\frac{\pi(\hbar v_F)^2 (n_h^0+\delta n)}{2T_0^2}\right)
\label{muH0}
\ee
where $G(y)$ is the inverse function of $F_1(x)$, 
\be 
y=F_1(x) \ \Leftrightarrow \ x=G (y)\equiv F_1^{-1}(y),
\ee
with $F_1$ defined in (\ref{funF}). 
Equations (\ref{muE})--(\ref{muH0}), together with (\ref{en-relax}) and (\ref{recombination}), give six equations for six unknown variables. The quantities $\omega$, $I$, $T_0$ and $\mu_0$ (or $E_F$), as well as parameters $\tau_\epsilon$ and $\tau_{\rm rec}$ are assumed to be known input parameters.

\subsection{Steady-state equations}

In the rest of the paper we will analyze the steady-state solutions of Eqs. (\ref{en-relax}), (\ref{recombination}). Setting $\p/\p t=0$ we get
\be 
I=\frac{{\cal E}(\mu_e,\mu_h,T)-{\cal E}(\mu_e^0,\mu_h^0,T_0)}{\tau_\epsilon A^{\rm intra}(\omega,\mu_e,\mu_h,T)} =
\frac {\hbar\omega\delta n}{\tau_{\rm rec}A^{\rm inter}(\omega,\mu_e,\mu_h,T)}
\left(1+\frac{\delta n}{n_e^0+n_h^0}\right).
\label{steadystate}
\ee
The second equation here does not contain the intensity $I$. It can be presented in the dimensionless form
\be 
{\cal L}(\hbar\omega,\mu_e,\mu_h,
T)=\frac{\tau_\epsilon}{\tau_{\rm rec}}
{\cal R}\left(\hbar\omega,\mu_e,\mu_h,
T,\Delta\right),
\label{L=R}
\ee
where the left- and right-hand sides read
\be 
{\cal L}(\hbar\omega,\mu_e,\mu_h,T)=
\frac{\textrm{Re }\sigma^{(1)}_{\rm inter}(\omega,\mu_e,\mu_h,T)}
{\textrm{Re }\sigma^{(1)}_{\rm intra}(\omega,\mu_e,\mu_h,E_i,T)}, \label{calL}
\ee
\be
{\cal R}\left(\hbar\omega,\mu_e,\mu_h,T,\Delta\right)
=
\frac{\Delta
\left(1+\Delta\right)\hbar\omega T_0^2\left[F_1\left(-\frac{\mu_0}{T_0}\right)+
F_1\left(\frac{\mu_0}{T_0}\right)\right] }
{T^3\left[F_2\left(\frac{\mu_e}T\right)
+F_2\left(-\frac{\mu_h}T\right)\right]
-T_0^3\left[F_2\left(\frac{\mu_e^0}{T_0}\right)
+F_2\left(-\frac{\mu_h^0}{T_0}\right)\right]
}.\label{calR}
\ee
Solving the nonlinear equation (\ref{L=R}), together with (\ref{muE})--(\ref{muH0}), we can find the relation between the relative change of the density $\Delta=\delta n/(n_e^0+n_h^0)$ and the relative change of the temperature $\delta T/T_0=(T-T_0)/T_0$. Notice that equation (\ref{L=R}) depends only on the ratio of the characteristic times $\tau_\epsilon/\tau_{\rm rec}$ and not on each of them separately. After the relation between $\Delta$ and $\delta T/T_0$ is found we substitute it into any of the equations (\ref{steadystate}) and relate these quantities to the intensity of radiation. This gives two equivalent formulas 
\be 
{\cal P}_a\equiv \frac{I}{I_0} = 
\frac{\hbar\omega}{T_0 }
\frac {\tau_\epsilon}{\tau_{\rm rec}}\frac{\Delta
\left(1+\Delta \right)}{A^{\rm inter}(\omega,\mu_e,\mu_h,T)} \left[F_1\left(\frac{\mu_0}{T_0}\right)+F_1\left(-\frac{\mu_0}{T_0}\right)\right],
\label{1a1}
\ee
or 
\be 
{\cal P}_b=\frac{I}{I_0}=
\frac{(T/T_0)^3\left[F_2\left(\frac{\mu_e}{T}\right)+F_2\left(-\frac{\mu_h}{T}\right)\right]- \left[F_2\left(\frac{\mu_e^0}{T_0}\right)+F_2\left(-\frac{\mu_h^0}{T_0}\right)\right]} {A^{\rm intra}(\omega,\mu_e,\mu_h,T)},
\label{2a1a}
\ee
where we have introduced a power density unit
\be 
I_0= \frac{2T_0^3}{\pi(\hbar v_F)^2\tau_\epsilon }\approx 
410 \frac{(T_0[\rm{K}]/300)^3 }{\tau_\epsilon[\textrm{ps}] } \frac{\textrm{W}}{\textrm{cm}^2},
\ee 
which does not depend on the charge carrier density. At room temperature the power density unit $I_0$ is about 400 W/cm$^2$ if $\tau_\epsilon\simeq 1$ ps. This is a rather small value, i.e., the strongly nonlinear regime corresponds to $I\gg I_0$. Notice that Eqs. (\ref{L=R})--(\ref{2a1a}) are presented in the explicitly dimensionless form, which means, in particular, that scaling all energies by the same numerical factor will not change the final results. This makes them universal in a sense.

Now we can start analyzing results which our model gives. We will assume that $T_0=300$ K and $\mu_0=-0.2$ eV which corresponds to the equilibrium charge carriers density of about $3\times 10^{12}$ cm$^{-2}$, see (\ref{dens-example}). All energy quantities will be given in eV.

\section{Results\label{sec:results}}

\subsection{Density-temperature diagrams\label{sec:dens-temp}}

\begin{figure}[ht]
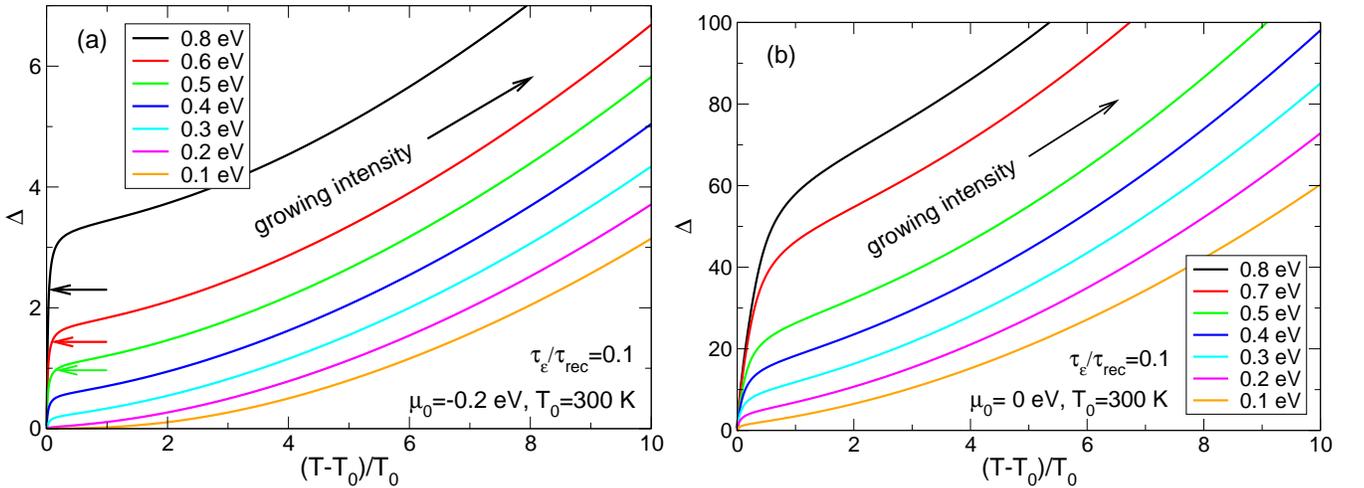

\includegraphics[width=0.49\textwidth]{fig2a.eps}
\includegraphics[width=0.49\textwidth]{fig2b.eps}
\caption{\label{fig:dndt}The relative change of the density $\Delta=\delta n/(n_e^0+n_h^0)$ versus relative change of the temperature $(T-T_0)/T_0$ at different values of the input wave frequency $\hbar \omega$ (in eV). The ratio of the relaxation times is $\tau_\epsilon/\tau_{\rm rec}=0.1$, temperature $T_0=300$ K, and the energy $E_i=30$ meV. The equilibrium chemical potential is (a) $\mu_0=-0.2$ eV and (b) $\mu_0=0$ eV. Arrows in (a) show the points where $I/I_0=10^5$. }
\end{figure}

Figure \ref{fig:dndt} shows the density-temperature diagrams obtained by solving equation (\ref{L=R}) at different frequencies, $T_0=300$ K, $\tau_\epsilon/\tau_{\rm rec}=0.1$, and two different values of the equilibrium chemical potenial $\mu_0=-0.2$ and $0$ eV. Each point on each curve corresponds to a certain value of the input wave power density. The low-intensity regime $I/I_0\ll 1$ corresponds to the origin of the plots where $\Delta\ll 1$ and $\delta T/T_0\ll 1$. The higher the power, the larger is the relative changes of both the density and the temperature, so that both $\Delta$ and $\delta T/T_0$ grow with the increasing intensity. The rate of their growth depends however on the frequency.

First, we consider the case $\mu_0=-0.2$ eV, Figure \ref{fig:dndt}(a). If $\hbar\omega\gtrsim 2|\mu_0|=0.4$ eV (black to blue curves), i.e., when the \textit{inter-band} transitions dominate, the relative change of the density $\Delta$, at low intensities, is much stronger than the relative change of the temperature $\delta T/T_0$. When $I/I_0$ grows from zero up to the values $\sim 10^5$ (see arrows in Figure \ref{fig:dndt}(a)), $\Delta$ increases up to $\Delta\simeq 1- 3$, while the temperature remains practically unchanged, $\delta T/T_0\ll 1$. The values $\Delta\simeq 1- 3$ correspond to practically equal densities of photo-excited electrons and holes; remind that at $\mu_0=-0.2$ eV and $T_0=300$ K the equilibrium electron and hole densities differed by almost five orders of magnitude, Eq. (\ref{dens-example}). At even larger intensities, $I/I_0\gtrsim 10^5$, the hot carrier temperature starts to grow too: in this situation the density of electrons and holes are large and close to each other and the both gases are heated by the intra-band absorption in the corresponding energy bands.

In the regime $\hbar\omega\ll 2|\mu_0|$ the tendency is opposite. The temperature $\delta T/T_0$ grows much faster than the relative change of the density $\Delta$ which is physically clear since at $\hbar\omega\ll 2|\mu_0|$ the \textit{intra-band} transitions dominate. At very large power densities, however, both $\Delta$ and $\delta T/T_0$ becomes quite comparable to each other.

If the case of intrinsic graphene, Figure \ref{fig:dndt}(b), the relative density changes are always much larger than the relative change of temperature. The absolute values of $\Delta$ are also much larger than in Figure \ref{fig:dndt}(a). These two features are a simple consequence of the fact that at $\mu_0=0$ eV the inter-band transitions are always dominant and that in equilibrium the density of charge carriers (\ref{dens-example-2}) is much smaller than in the doped one. For example, $\Delta\simeq 100$ in Figure \ref{fig:dndt}(b) corresponds to approximately the same values of the non-equlibrium electron and holes densities ($n_e=n_h\approx 1.63\times 10^{13}$ cm$^{-2}$) as $\Delta\simeq 4$ in Figure \ref{fig:dndt}(a) ($n_e\approx 1.25\times 10^{13}$ cm$^{-2}$, $n_h\approx 1.56\times 10^{13}$ cm$^{-2}$).

\begin{figure*}[ht]
\includegraphics[width=\textwidth]{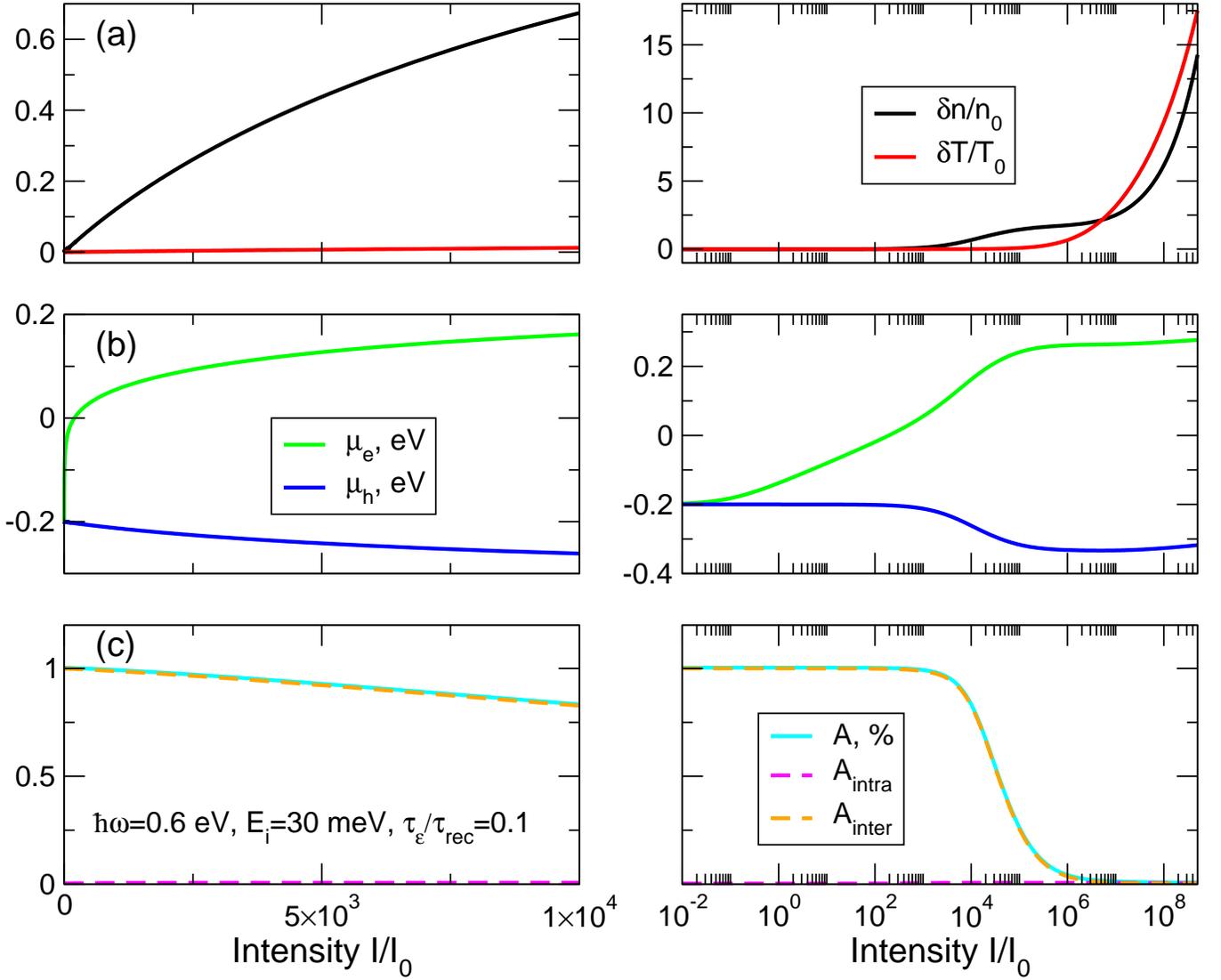}
\caption{\label{fig:doped0.6} (a) The relative change of the density $\Delta=\delta n/n_0=\delta n/(n_e^0+n_h^0)$ and temperature $(T-T_0)/T_0$, (b) the chemical potentials of electrons and holes $\mu_e$ and $\mu_h$, and (c) the absorption coefficient $A$, together with the intra- and inter-band contributions, as functions of the dimensionless power density $I/I_0$ at $\hbar\omega= 0.6$ eV, $E_i=30$ meV, and $\tau_\epsilon/\tau_{\rm rec}=0.1$. The equilibrium chemical potential and temperature are $\mu_0=-0.2$ eV and $T_0=300$ K. Left columns show the power dependencies at $I/I_0<10^4$, the right columns -- the same dependencies in a broader power range. The horizontal axis scale is the same in all left and all right panels. The vertical axis scale is the same in both (c) panels. }
\end{figure*}

\subsection{Power dependencies of different physical quantities\label{sec:power-depend}}

Now we consider how different physical quantities vary with the radiation intensity. We show results for doped ($\mu_0=-0.2$ eV) and intrinsic graphene ($\mu_0=0$ eV).

\subsubsection{Doped graphene}

Figure \ref{fig:doped0.6} exhibits the power dependencies of the chemical potentials $\mu_e$ and $\mu_h$, temperature $T$, photo-excited charge carrier density $\delta n$ and the absorption coefficient $A$, in the doped graphene sample with $\mu_0=-0.2$ eV, under the condition $\hbar\omega>2|\mu_0|$ when the inter-band transitions dominate. The ratio of relaxation times is assumed to be $\tau_\epsilon/\tau_{\rm rec}=0.1$ and we consider a relatively high-mobility sample with $E_i=30$ meV (this corresponds to $N_i\approx 6.4\times 10^{11}$ cm$^{-2}$ and $\mu\approx 7260$ cm$^2$/Vs, see Fig. \ref{fig:param}). At low intensities $I/I_0\lesssim 10^4$, left panels, the density $\delta n$ substantially changes, as expected, while the temperature remains practically unchanged. The chemical potential of electrons quickly grows, from the initial value $\mu_0=-0.2$ eV, and becomes positive at $I/I_0\simeq 200$; the chemical potential of holes becomes more negative and varies slowly. The absorption coefficient does not change with the intensity up to $I/I_0\simeq 10^3$, but starts to decrease when $I/I_0$ approaches the values of order of $10^4$. It is mainly due to the inter-band contribution, and is about 1\%. This number differs from traditional 2.3\% since we consider graphene lying on a SiO$_2$ substrate, Eqs. (\ref{absorptn-intra})--(\ref{absorptn-inter}).

When the intensity grows further, electron and holes gases get heated and the charge carrier temperature increases too. The relative changes of density and temperature becomes equal at $I/I_0\approx 5.4\times 10^6$, Figure \ref{fig:doped0.6}(a), right panel. At the point, where the black and red curves intersect, $\Delta=\delta T/T_0\approx 2.14$, which corresponds to $n_e\approx 6.68\times 10^{12}$ cm$^{-2}$, $n_h\approx 9.80\times 10^{12}$ cm$^{-2}$, and $T\approx 3.14 T_0 \approx 942$ K. The chemical potential of electrons continues to grow reaching the values $\simeq +0.26$ eV at $I/I_0\simeq 10^6$. The chemical potential of holes gets more negative and becomes equal to  $\simeq -0.33$ eV at $I/I_0\simeq 10^6$. Since the occupation of electrons and holes states around $\hbar\omega/2\simeq 0.3$ becomes much less asymmetric as compared to equilibrium the inter-band absorption starts to fall down at $I/I_0\simeq 10^4$ and becomes about $0.04$\% at $I/I_0\simeq 10^6$. This strong reduction of the absorption coefficient is usually referred to as the saturable absorption effect. The intra-band absorption remains small as compared to the inter-band one. 

\begin{figure*}[ht]
\includegraphics[width=\textwidth]{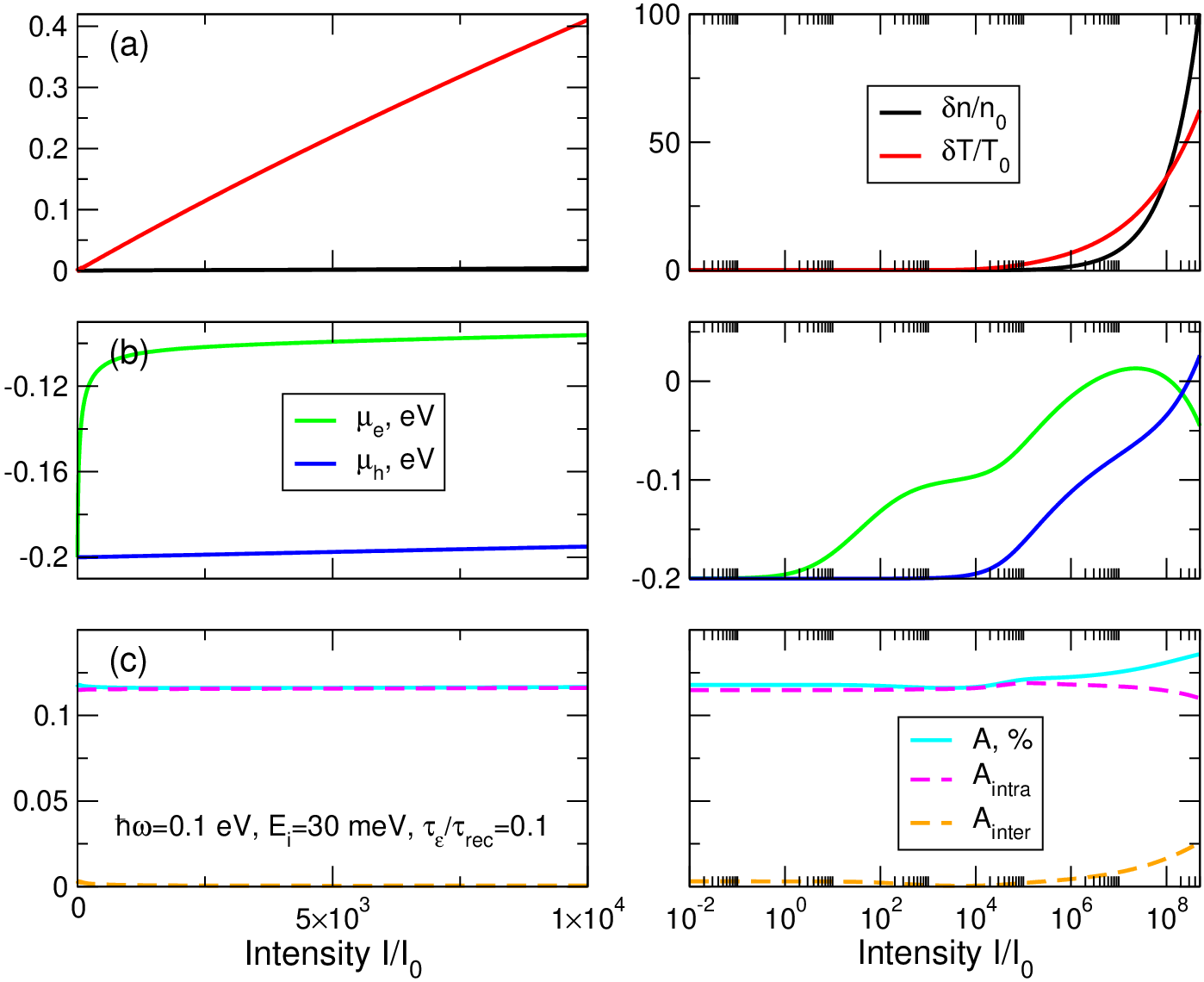}
\caption{\label{fig:doped0.1} (a) The relative change of the density $\Delta=\delta n/n_0=\delta n/(n_e^0+n_h^0)$ and temperature $(T-T_0)/T_0$, (b) the chemical potentials of electrons and holes $\mu_e$ and $\mu_h$, and (c) the absorption coefficient $A$, together with the intra- and inter-band contributions, as functions of the dimensionless power density $I/I_0$ at $\hbar\omega= 0.1$ eV, $E_i=30$ meV, and $\tau_\epsilon/\tau_{\rm rec}=0.1$. The equilibrium chemical potential and temperature are $\mu_0=-0.2$ eV and $T_0=300$ K. Left columns show the power dependencies at $I/I_0<10^4$, the right columns -- the same dependencies in a broader power range. The horizontal axis scale is the same in all left and all right panels. The vertical axis scale is the same in both (c) panels. }
\end{figure*}

Now we consider the case $\hbar\omega<2|\mu_0|$ where the intra-band absorption plays the crucial role at low intensities. Figure \ref{fig:doped0.1} shows the power dependencies of different physical quantities under the same conditions as in Figure \ref{fig:doped0.6} but at $\hbar\omega=0.1$ eV. Now the temperature substantially grows at low intensities while the charge carrier density remains almost unchanged up to $I/I_0\simeq 10^5$, Figure \ref{fig:doped0.1}(a), right panel. The chemical potential of electrons sharply grows at the radiation power $I/I_0\lesssim 700$ but then saturates at the much lower level $\mu_e\simeq -0.1$ eV than in Figure \ref{fig:doped0.6}. The chemical potential of holes remains almost constant slightly decreasing in the absolute value. The absorption coefficient is about 0.12\% at low intensities and remains practically constant up to $I/I_0\simeq 10^5$, Figure \ref{fig:doped0.1}(c). It is mainly due to the intra-band contribution which is much smaller than in the previous example since the frequency lies in the gap between the intra- and inter-band absorption areas, $\gamma(\mu_0)\ll \omega\ll 2|\mu_0|/\hbar$, where $\hbar\gamma(\mu_0)\approx 4.5$ meV under our conditions. At $I/I_0\gtrsim 10^5$ the absorption coefficient starts to grow (the induced absorption \cite{Winnerl11}), mainly due to the inter-band contribution which becomes essential since the occupation of energy levels at $E\simeq -\hbar\omega/2$ is no longer negligible due to the heating of the hole gas. At $I/I_0\gtrsim 10^5$ also the chemical potentials of both electrons and holes start to substantially grow making the distribution of charge carriers over the bands more uniform; at $I/I_0$ between $\simeq 10^6$ and $10^8$ the chemical potential of electrons even becomes positive, Figure \ref{fig:doped0.1}(b), right panel. At even larger intensities $I/I_0\gtrsim 10^8$ the chemical potential of holes moves to the conduction band, $\mu_h>0$, while that of electrons becomes negative again, Figure \ref{fig:doped0.1}(b), right panel.

\begin{figure}[ht]
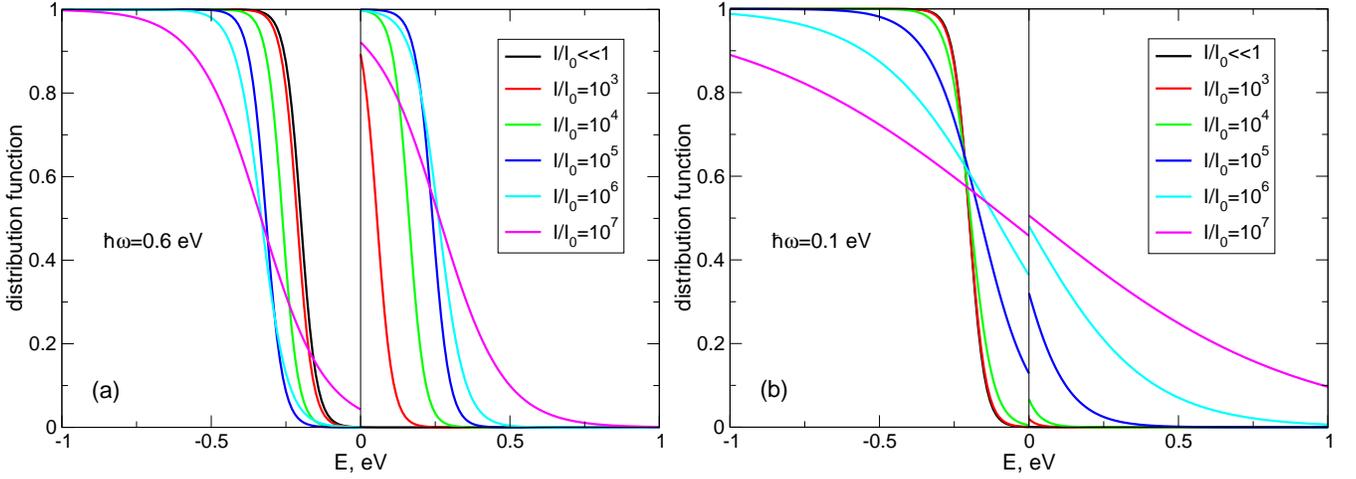

\includegraphics[width=0.49\textwidth]{fig5a.eps}
\includegraphics[width=0.49\textwidth]{fig5b.eps}
\caption{\label{fig:Fermi_doped} The electron distribution function (\ref{f_glob}) in the valence ($E<0$) and conduction ($E>0$) bands at different values of the radiation intensity. The radiation frequency is (a) $\hbar\omega=0.6$ eV and (b) $\hbar\omega=0.1$ eV, other parameters are: $\mu_0=-0.2$ eV, $T_0=300$ K, $E_i=30$ meV, and $\tau_\epsilon/\tau_{\rm rec}=0.1$. } 
\end{figure}

In Figure \ref{fig:Fermi_doped} we further illustrate our results by showing the electron distribution function in the valence and conduction bands at different radiation intensities. Here one clearly sees a qualitative difference between the charge carrier distributions at $\hbar\omega>2|\mu_0|$, Figure \ref{fig:Fermi_doped}(a), and at $\hbar\omega<2|\mu_0|$, Figure \ref{fig:Fermi_doped}(b). In the inter-band absorption case $\hbar\omega>2|\mu_0|$ the slope of the $f(E)$ curves, and hence the charge carrier temperature $T$, remains practically unchanged up to intensities $I/I_0\simeq 10^5$. In contrast, the chemical potentials vary quite strongly: already at $I/I_0\simeq 10^3$ (red curve) the chemical potential of electrons is positive and the occupation of the conduction band is quite large. At higher intensities, $I/I_0\gtrsim 10^6$, when the densities of photo-excited electrons and holes become comparable, the temperature starts to grow too due to the intra-band absorption in each band, and the slope of the $f(E)$ curves decreases. 

In the intra-band absorption case $\hbar\omega<2|\mu_0|$, Figure \ref{fig:Fermi_doped}(b), the slope of the curves noticeably decreases already at $I/I_0\gtrsim 10^4$, and the occupation of the conduction band is much weaker than in the previous case (the chemical potential of electrons remains negative). Only at the very high intensity $I/I_0\simeq 10^7$ (magenta curve) $\mu_e$ becomes slightly positive.

\subsubsection{Intrinsic graphene}

\begin{figure*}[ht]
\includegraphics[width=\textwidth]{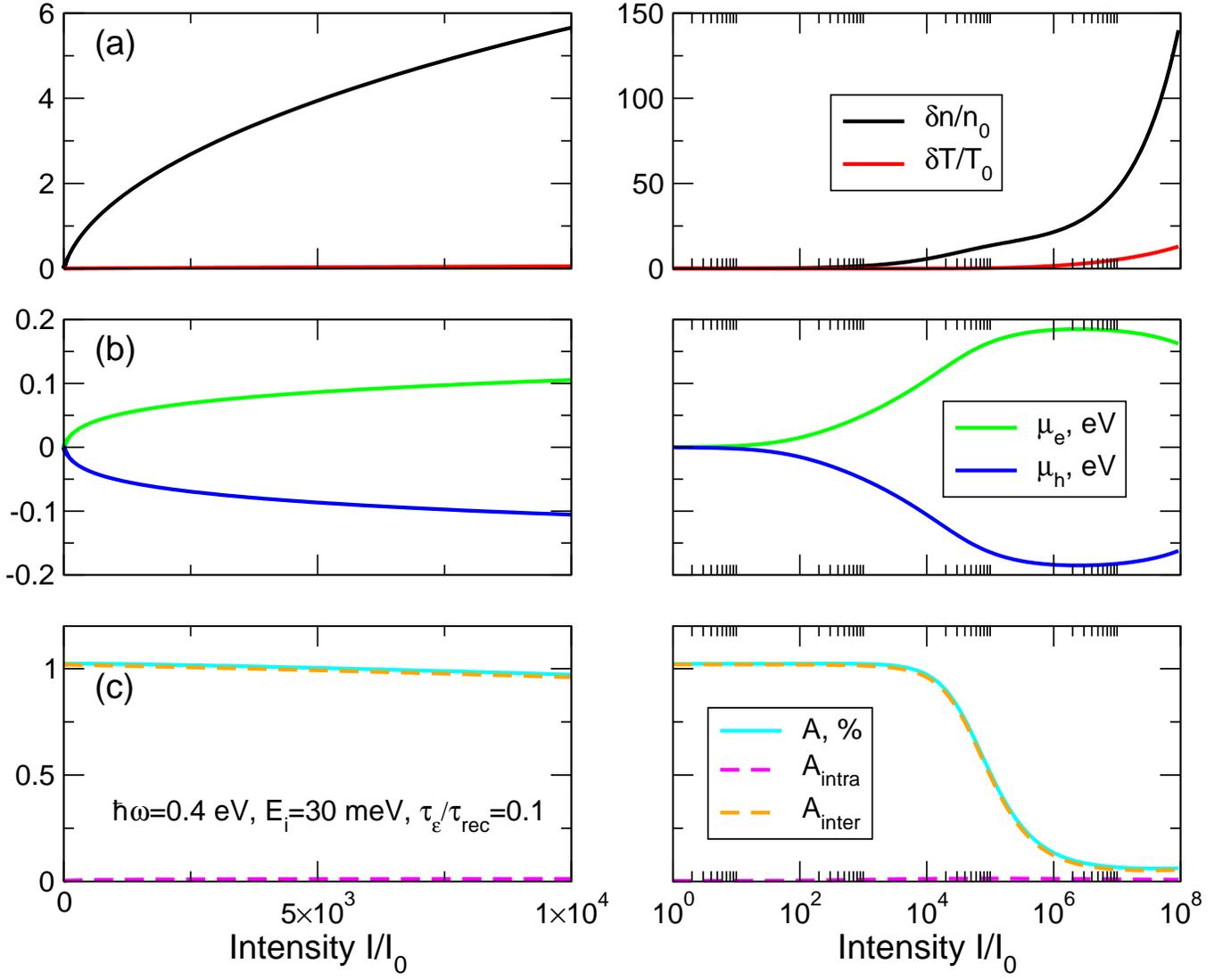}
\caption{\label{fig:w0.4Ei30TT0.1mu0} (a) The relative change of the density $\Delta=\delta n/n_0=\delta n/(n_e^0+n_h^0)$ and temperature $(T-T_0)/T_0$, (b) the chemical potentials of electrons and holes $\mu_e$ and $\mu_h$, and (c) the absorption coefficient $A$, together with the intra- and inter-band contributions, as functions of the dimensionless power density $I/I_0$ at $\hbar\omega= 0.4$ eV, $E_i=30$ meV, and $\tau_\epsilon/\tau_{\rm rec}=0.1$. The equilibrium chemical potential and temperature are $\mu_0=0$ eV (intrinsic graphene) and $T_0=300$ K. Left columns show the power dependencies at $I/I_0<10^4$, the right columns -- the same dependencies in a broader power range. The horizontal axis scale is the same in all left and all right panels. The vertical axis scale is the same in left and right (b) and (c) panels. } 
\end{figure*}

Figure \ref{fig:w0.4Ei30TT0.1mu0} shows the power dependencies of different physical quantities in intrinsic graphene with $\mu_0=0$ and $\hbar\omega=0.4$ eV. Other parameters ($T_0$, $E_i$ and $\tau_\epsilon/\tau_{\rm rec}$) are the same as in Figures \ref{fig:doped0.6}, \ref{fig:doped0.1} and \ref{fig:Fermi_doped}. Since at $\mu_0=0$ the condition $\hbar\omega>2|\mu_0|$ is always satisfied the curves shown in Figure \ref{fig:w0.4Ei30TT0.1mu0} are qualitatively similar to those from Figure \ref{fig:doped0.6}. The inter-band transitions dominate, therefore the relative change of the density $\Delta$ is always much larger than the relative change of temperature, Figure \ref{fig:w0.4Ei30TT0.1mu0}(a). Only at $I/I_0\gtrsim 10^6$ the temperature $T$ starts to noticeably grow, Figure \ref{fig:w0.4Ei30TT0.1mu0}(a), right panel. The chemical potentials of electrons and holes are symmetric, $\mu_e=-\mu_h$, and achieve the values of order $|\mu_e|=|\mu_h|\simeq 0.1$ eV at $I/I_0\gtrsim 10^4$ and $\simeq 0.2$ eV at $I/I_0\gtrsim 10^5-10^7$, Figure \ref{fig:w0.4Ei30TT0.1mu0}(b). The absorption curves show the saturable absorption effect: the absorption coefficient $A$ falls down from the value $\simeq 1$ \% at $I/I_0\lesssim 10^4$ down to the values $\lesssim 0.12$ \% at $I/I_0\gtrsim 10^6$. The contribution of the intra-band absorption to $A$ is negligibly small.

\begin{figure}[ht]
\includegraphics[width=0.49\textwidth]{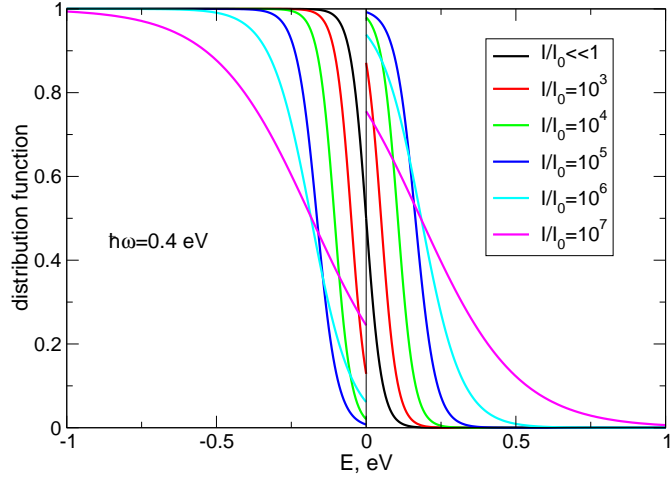}
\caption{\label{fig:Fermi_intrinsic} The electron distribution function (\ref{f_glob}) in the valence ($E<0$) and conduction ($E>0$) bands of intrinsic graphene ($\mu_0=0$ eV) at different values of the radiation intensity. Other parameters are: $\hbar\omega=0.4$ eV, $T_0=300$ K, $E_i=30$ meV, and $\tau_\epsilon/\tau_{\rm rec}=0.1$. } 
\end{figure}

Figure \ref{fig:Fermi_intrinsic} shows the non-equilibrium distribution function of electrons and holes at $\hbar\omega=0.4$ eV and different power levels. In accordance with Figure \ref{fig:w0.4Ei30TT0.1mu0} the temperature remains low ($T\simeq T_0$) at the intensities up to $I/I_0\simeq 10^4$, while the chemical potentials $|\mu_{e,h}|$ grow. At higher intensities the temperature increases too (the cyan and, especially, magenta curves).

\subsection{Frequency dependencies of different physical quantities \label{sec:absorption}}

Another way to clarify the physics of the discussed phenomena is to analyze how the frequency dependencies of the absorption coefficient $A(\omega)$ and other physical quantites are modified under the influence of the strong radiation power. This is especially important question since the spectra $A(\omega)$ can be directly experimentally measured \cite{Li08}. 

\begin{figure}[ht]
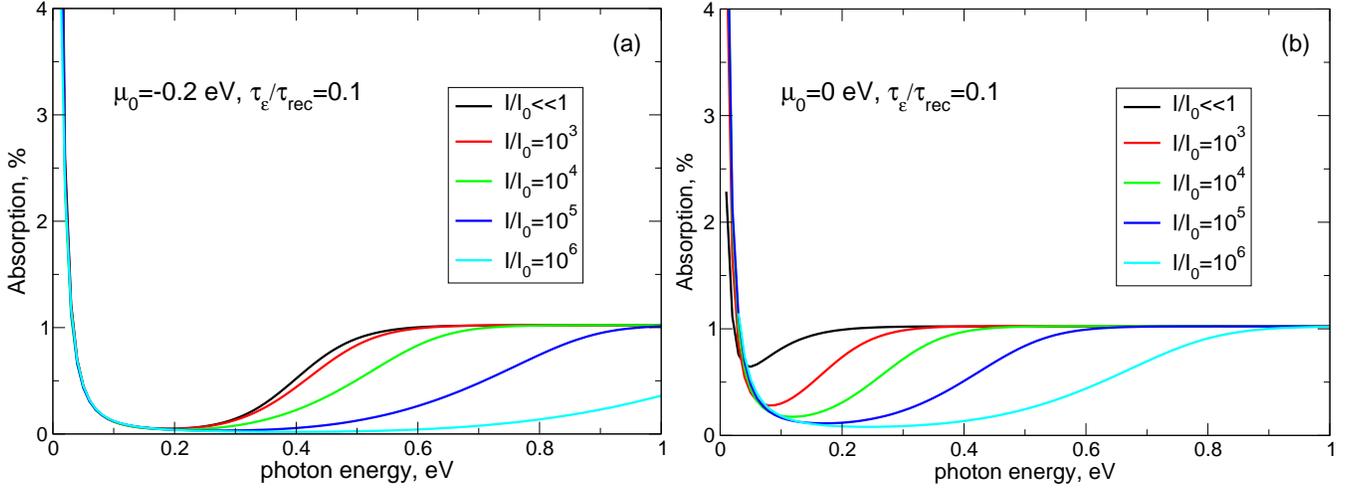

\includegraphics[width=0.49\textwidth]{fig8a.eps}
\includegraphics[width=0.49\textwidth]{fig8b.eps}
\caption{\label{fig:Absorp} The absorption coefficient vs frequency at different values of the radiation intensity, for 
(a) $|\mu_0|=0.2$ eV and (b) $\mu_0=0$ eV. Other parameters are  $\tau_\epsilon/\tau_{\rm rec}=0.1$, $E_i=30$ meV and $T_0=300$ K. } 
\end{figure}

Figure \ref{fig:Absorp} shows the power-dependent absorption spectra in (a) doped ($|\mu_0|=0.2$ eV) and (b) intrinsic ($\mu_0=0$ eV) graphene, at $\tau_\epsilon/\tau_{\rm rec}=0.1$, $E_i=30$ meV and $T_0=300$ K. The black curves in both panels exhibit the known linear response absorption curves (e.g., Refs. \cite{Mikhailov16a,Li08}) corresponding to the equilibrium chemical potential, $\mu_e=\mu_h=\mu_0$, and equilibrium temperature, $T=T_0$. One sees that the growing power substantially reduces the absorption in graphene. This \textit{saturable absorption} effect is the case already at $\hbar\omega\gtrsim |\mu_0|$ (not $2|\mu_0|$!) in Figure \ref{fig:Absorp}(a) and at $\hbar\omega\gtrsim 0.04$ eV in Figure \ref{fig:Absorp}(b). The influence of the radiation power is noticeable at $I/I_0\simeq 10^3$ in the doped graphene, Figure \ref{fig:Absorp}(a), and at even lower powers ($I/I_0\simeq 10^1-10^2$) in the intrinsic graphene, Figure \ref{fig:Absorp}(b). Quantitatively, the suppression of $A(\omega)$ is very strong; for example, in doped graphene at $\hbar\omega=0.6$ eV and $I/I_0= 10^6$ the absorption is only 0.042\%, i.e. it is reduced by a factor of $\sim 24$. In the intrinsic graphene at $\hbar\omega=0.3$ eV and $I/I_0= 10^6$ the absorption is reduced down to 0.085\%, i.e. by a factor of $\sim 12$. The saturable absorption effect at high frequencies ($\hbar\omega \gtrsim |\mu_0|$) was experimentally observed in many experiments, see, e.g., Refs. \cite{Bao09,Zhang09,Winnerl11,Zheng12,Bianchi17}. 

At lower frequencies ($\hbar\omega \lesssim |\mu_0|$) our model predicts an essentially different behavior. The absorption spectrum weakly depends on the radiation power at $\hbar\omega \lesssim |\mu_0|$ and the radiation may lead to a \textit{slight increase} of the absorption. In doped graphene, Figure \ref{fig:Absorp}(a), this effect is rather small; for example, at $\hbar\omega=0.1$ eV the absorption coefficient is about 11.79\% at $I/I_0\ll 1$ and increases by $\simeq 0.3$\% at $I/I_0=10^5$ and by $\simeq 0.45$\% at $I/I_0=10^6$. In intrinsic graphene, Figure \ref{fig:Absorp}(b), the growth of absorption is stronger and can achieve $2-3$\% at $I/I_0$ up to $\simeq 10^4$: for example, at $\hbar\omega=0.01$ eV the absorption is about 2.3\% at $I/I_0\ll 1$, 4.8\% at $I/I_0= 10^3$ and 5.8\% at $I/I_0= 10^4$. The specific numbers of the absorption change depend of course on the chosen parameters of the structure.

\begin{figure}[ht]
\includegraphics[width=0.32\columnwidth]{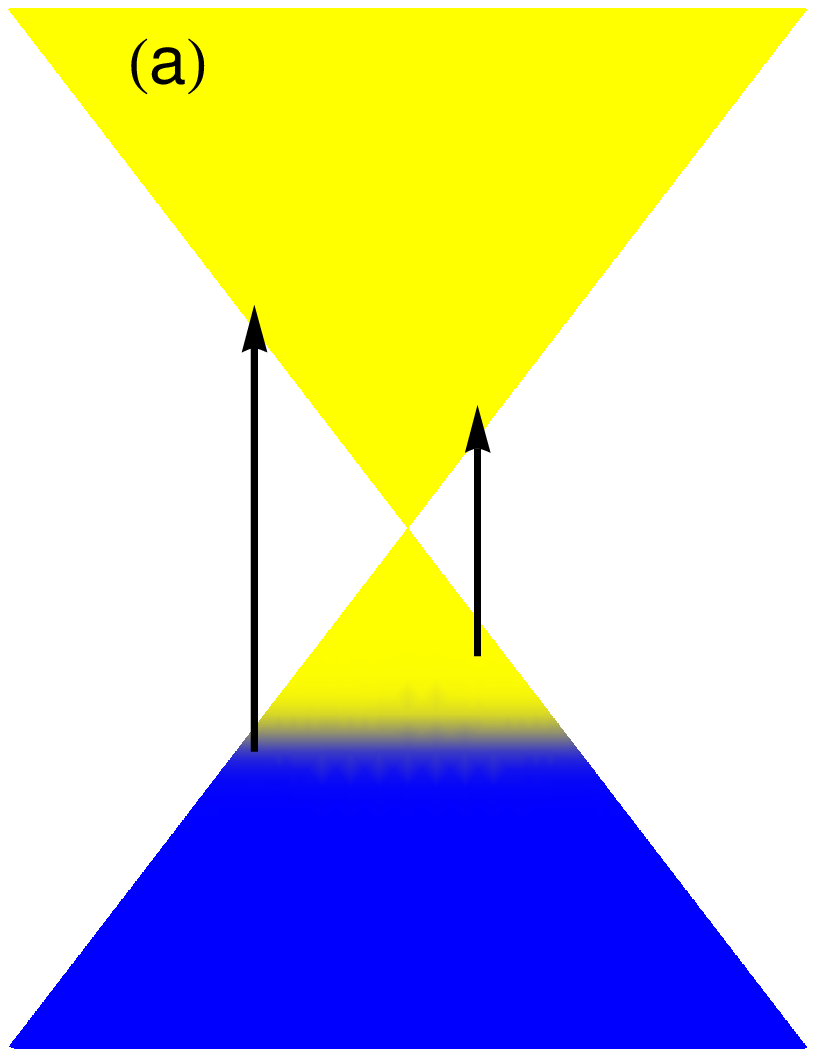} 
\includegraphics[width=0.32\columnwidth]{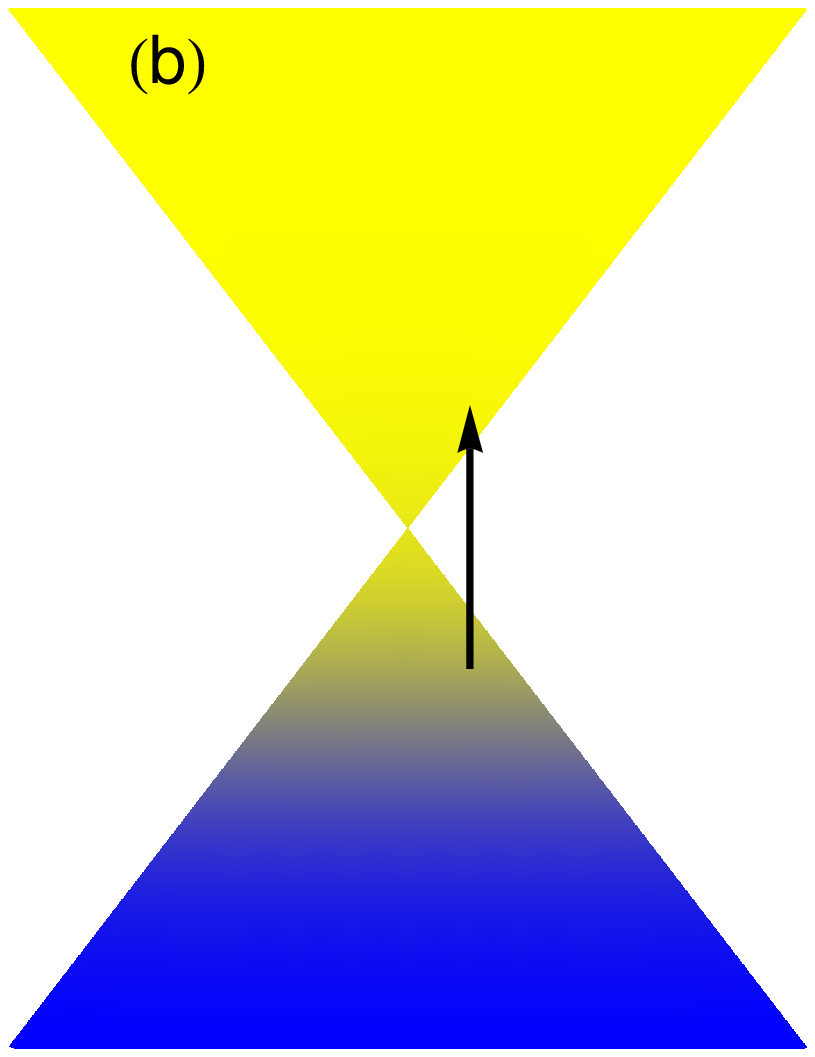} 
\includegraphics[width=0.32\columnwidth]{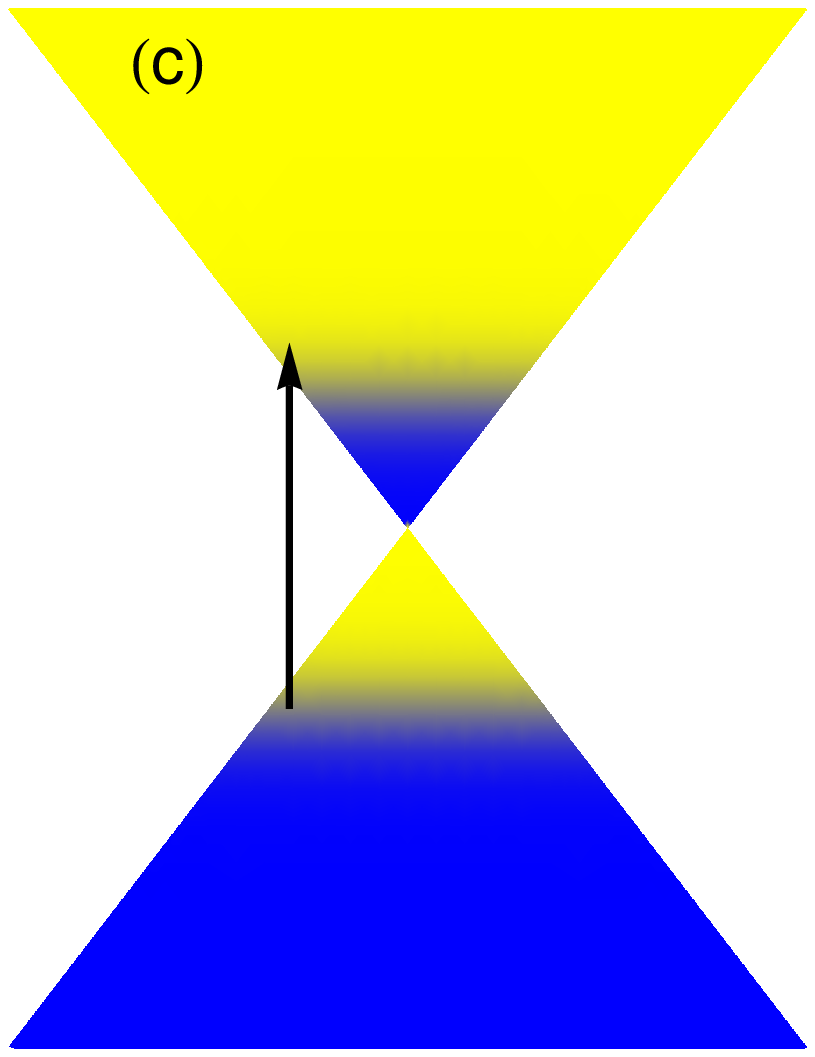}
\caption{\label{fig:transitions} Inter-band transitions in graphene (a) in equilibrium and under (b) low-frequency and (c) high-frequency powerful irradiation at a negative chemical potential $\mu_0<0$ and low temperature $T_0$. (a) In equilibrium the high-frequency photons are absorbed (left arrow: electrons jump from occupied initial to the empty final states) and the low-frequency photons are not (right arrow: initial states are empty). (b) Under the low-frequency powerful irradiation the electron gas gets heated due to the intra-band absorption, $T_0\to T$, the initial states in the valence band get partly occupied and the inter-band absorption increases (induced absorption). (c) Under the high-frequency powerful irradiation the final states in the conduction band get partly occupied and the inter-band absorption decreases (absorption saturation).} 
\end{figure}

\begin{figure}[ht]
\includegraphics[width=0.49\textwidth]{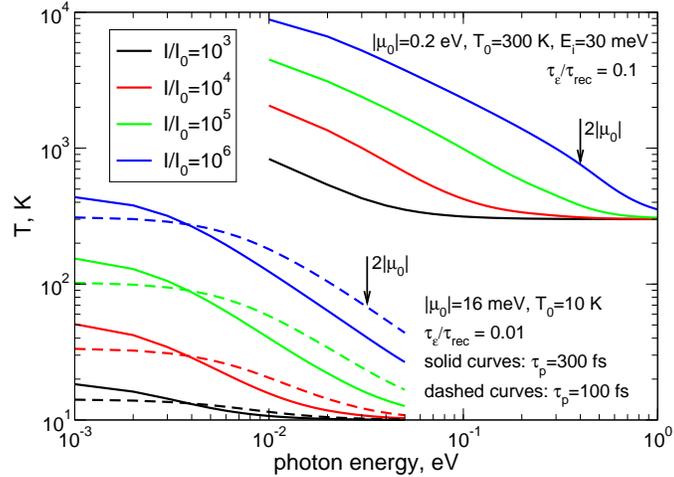} 
\caption{\label{fig:TvsW} The hot electron temperature $T$ as a function of the photon energy. Four curves in the upper right corner are plotted for parameters corresponding to Figure \ref{fig:Absorp}(a). The curves in the lower left corner are plotted for parameters of the experiment \cite{Winnerl11}, see discussion in the text. Arrows indicate the positions of the double chemical potential $2|\mu_0|$.  } 
\end{figure}

Physically the growth of absorption at $\hbar\omega\lesssim 2|\mu_0|$ and its reduction at $\hbar\omega\gtrsim 2|\mu_0|$ are explained by the radiation induced redistribution of electrons over quantum states in the conduction and valence bands, Figure \ref{fig:transitions}. Notice that this qualitative picture, which appeared in many publications (see, e.g., Ref. \cite{Winnerl11}), implies that the hot electron temperature $T$, as well as the chemical potentials of electrons and holes $\mu_e$ and $\mu_h$, essentially depend on the photon energy. Within our HEM we can quantitatively evaluate these dependencies. In Figure \ref{fig:TvsW} we plot the hot electron temperature $T$ as a function of the photon energy for a few sets of experimental parameters. The four curves in the upper right corner show the dependencies $T(\omega)$ for parameters corresponding to Figure \ref{fig:Absorp}(a) ($|\mu_0|=0.2$ eV, $T_0=300$ K, $E_i=30$ meV and $\tau_\epsilon/\tau_{\rm rec}=0.1$) and four different power levels. The arrow labeled as $2|\mu_0|$ indicates the position of the double chemical potential. One sees that at $\hbar\omega\gtrsim 2|\mu_0|$ all curves tend to the equilibrium temperature value $T\to T_0=300$ K. Even if at low frequencies the hot electron temperature exceeds $T_0$ by more than one order of magnitude, at $\hbar\omega\simeq 2|\mu_0|$ it is already almost equal to $T_0$; for example, for $I/I_0=10^5$ (green curve) the temperature $T$ drops from 4486 K at $\hbar\omega=10$ meV down to 380 K at $\hbar\omega=2|\mu_0|=0.4$ eV.

In a recent experiment on multilayer epitaxial graphene \cite{Winnerl11} the influence of a strong (pump) radiation on the transmission coefficient of the weak (probe) wave was studied, and a few-percent increase (decrease) of absorption was observed at $\hbar\omega \lesssim 2|\mu_0|$ ($\hbar\omega \gtrsim 2|\mu_0|$). Graphene was weakly doped ($\mu_0$ was evaluated to be $\simeq 13$ meV) and the experiment was performed at 10 K. The pump radiation with the fluence up to $\simeq 1$ $\mu$J/cm$^2$ reduced the absorption coefficient by a few percent at $\hbar\omega=30$ meV and increased it by a few percent at $\hbar\omega=20$ meV. The authors interpreted their results applying a \textit{simplified} HEM which assumed that the chemical potentials remain unchanged under the action of radiation, $\mu_e=\mu_h=\mu_0$,  and the hot electron temperature does not depend on the radiation frequency, $T(\omega)=$const. A similar effect was observed and the same interpretation was applied to its explanation in Ref. \cite{Alexander18}, where the nonlinear absorption in graphene was measured under different equilibrium conditions with $\mu_0=-0.4$ eV and $T_0=300$ K.

\begin{figure}[ht]
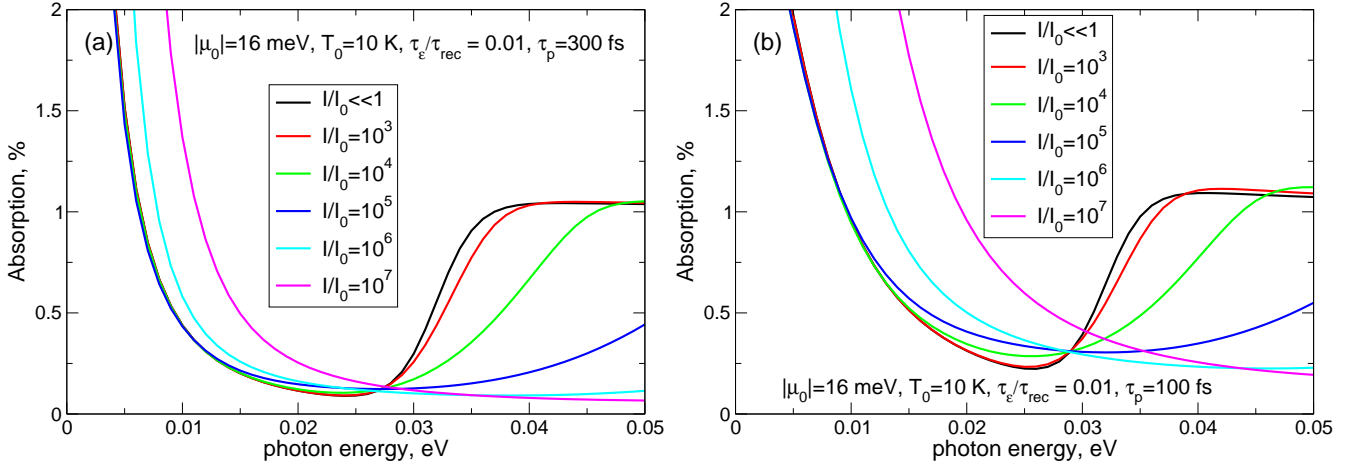

\includegraphics[width=0.49\textwidth]{fig11a.eps}
\includegraphics[width=0.49\textwidth]{fig11b.eps}
\caption{\label{fig:AbsorpT} (a) The absorption coefficient $A$ as a function of the radiation frequency at $\tau_\epsilon/\tau_{\rm rec}=0.01$, $|\mu_0|=16$ meV, $T_0=10$ K, and (a) $\tau_p=300$ fs and (b) $\tau_p=100$ fs. } 
\end{figure}

As we have seen above, in reality the quantities $T$, $\mu_e$, and $\mu_h$ cannot be considered as frequency independent and a more general theory should be applied. In Figure \ref{fig:AbsorpT}(a) we plot the absorption spectra calculated for $|\mu_0|=16$ meV, $T_0=10$ K, $\tau_p=300$ fs, and $\tau_\epsilon/\tau_{\rm rec}=0.01$; the results were found to be weakly dependent on $\tau_\epsilon/\tau_{\rm rec}$ as long as this parameter is small as compared to unity. To calculate the graphene conductivity and the absorption coefficient we used the model of the energy independent momentum scattering time (\ref{GammaEindep}) with $\tau_p$ taken from Ref. \cite{Winnerl11}. One sees that the curves corresponding to different power levels intersect at one point lying approximately at $\hbar\omega\approx 28$ meV. This value is smaller than $2|\mu_0|=32$ meV (in contrast to the results of the simplified HEM, see Ref. \cite{Winnerl11}); this difference results from strong frequency dependence of $T(\omega)$, shown in the lower left corner of Figure \ref{fig:TvsW}. Right and left from 28 meV the absorption decreases and increases respectively, and we have chosen the chemical potential $|\mu_0|=16$ meV to get approximately equal (in absolute values) changes of the absorption coefficient at $\hbar\omega= 20$ and $30$ meV. This quantity is slightly larger than $|\mu_0|=13$ meV extracted in Ref. \cite{Winnerl11} from the comparison of experimental data with the simplified HEM; the reason is again due to the frequency dependent (independent) hot electron temperature in the full (simplified) HEM. In general, one sees that our HEM gives reasonable results which can be used for analysis of different nonlinear optics experiments.

As seen from Figure \ref{fig:AbsorpT}(a) the reduction of absorption right from the intersection point (28 meV) is in general stronger than its growth left from this point. Qualitatively this is also in agreement with the experimental results of Ref. \cite{Winnerl11} (see Figures 3(a),(b) there). Physically the increase of absorption at low frequencies is due to the larger intra-band contribution to $A$. This contribution can be increased in samples with a lower mobility: in graphene layers with a smaller values of $\tau_p$, i.e., with a lower mobility, the hot electron temperature and the absorption increase at low frequencies should be larger. This is confirmed indeed in Figure \ref{fig:AbsorpT}(b) which show the absorption spectra for the same parameters as in Figure \ref{fig:AbsorpT}(a) but for three times smaller $\tau_p=100$ fs. One sees that the absorption change increase by several times at $\hbar\omega=20$ meV while at $\hbar\omega\gtrsim 30$ meV changes are less dramatic. Figure \ref{fig:TvsW} also confirms that at frequencies around $2|\mu_0|$ the temperature $T$ is larger in samples with $\tau_p=100$ fs than in those with $\tau_p=300$ fs.

\section{Summary\label{sec:conclusion}}

Experiments on the nonlinear graphene optics are very often performed at so strong excitation powers that the perturbation theory fails to adequately describe their results. The hot electron model presented in this paper allows to calculate the most important parameters of highly non-equilibrium charge carriers in graphene -- the chemical potentials of electrons and holes, as well as their effective temperature -- thus enabling to correctly describe its response to the powerful electromagnetic radiation. The model is physically transparent and employs essentially one fitting parameter -- the ratio $\tau_\epsilon/\tau_{\rm rec}$ of the intra-band energy relaxation time $\tau_\epsilon$ to the inter-band recombination time $\tau_{\rm rec}$. The derived system of strongly nonlinear differential equations (\ref{en-relax}), (\ref{recombination})--(\ref{muH0}) allows to calculate all physical quantities characterizing the nonlinear graphene response as functions of the incident wave frequency and power, equilibrium temperature, doping level, sample mobility, dielectric environment and so on. The developed theory reasonably describes available experimental data. Together with the already published perturbative theories of the nonlinear graphene response \cite{Cheng14a,Cheng15,Mikhailov16a} the work done here paves the way to a more accurate interpretation of nonlinear optics experiments and to the development of new optoelectronic devices for visible, infrared and terahertz spectral ranges. 

\begin{acknowledgments}
This work has received funding from the European Union's Horizon
2020 research and innovation programme Graphene Core 2 under Grant Agreement No.  785219.
\end{acknowledgments}

\appendix 

\section{A model for the intra-band scattering rate\label{app:gamma-intra}}

In order to calculate the intra-band conductivity we need a model for the energy dependent intra-band momentum relaxation rate $\gamma_{\rm intra}(E)$. It is known \cite{DasSarma11} that the most important scattering mechanism of electrons in graphene is the charged impurity scattering, and that at high energies $\gamma_{\rm intra}(E)$ is proportional to $N_i/|E|$, where $N_i$ is the impurity density. This can be written in the form\cite{Trushin07,Hwang09}  
\be 
\hbar\gamma_{\rm intra}(E)=
\frac{E_{i}^2}{|E|}, \ \ \ |E|\gtrsim E_i, \label{gam-large-E}
\ee
where 
\be 
E_i=\alpha\frac{e^2}{\kappa}\sqrt{\pi N_i}\label{coul-en-1}
\ee
is the Coulomb energy associated with the impurity density $N_i$, $\kappa$ is the effective dielectric constant of the medium surrounding the graphene layer, and $\alpha$ is a number of order unity. In Ref. \cite{Hwang09} the formula $\alpha=2\sqrt{I_0(r_s)}$ was derived, where 
\be 
r_s=\frac{e^2}{\kappa \hbar v_F},
\ee
is the effective fine structure constant of graphene and  
\be 
I_0(r_s)=\int_0^1dx\frac{x^2\sqrt{1-x^2}}{(x+2r_s)^2}.
\ee
If graphene lies on the surface of silicon dioxide, $\kappa_{{\rm SiO}_2}=3.9$, the effective dielectric constant is $\kappa=(\kappa_{{\rm SiO}_2}+1)/2=2.45$ and $r_s\approx 0.896$. Then $\alpha \approx 0.36$. 

\begin{figure}[ht]
\includegraphics[width=0.48\textwidth]{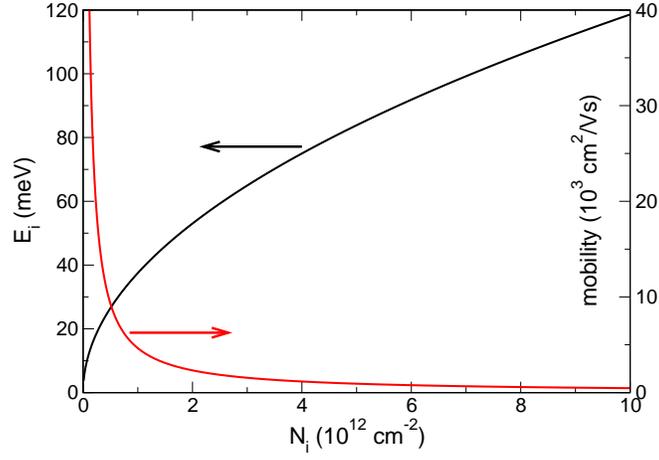}
\caption{\label{fig:param}The Coulomb impurity energy $E_i$ (in meV) and the mobility $\mu$ as a function of the impurity density $N_i$. Graphene is assumed to lie on a SiO$_2$ substrate so that $\kappa=2.45$, $\alpha=0.36$ and $\beta=9.61$. }
\end{figure}

The formula (\ref{gam-large-E}) is not valid at small energies (below $E_i$). We assume that it can be generalized as follows
\be 
\hbar\gamma_{\rm intra}(E)=\frac \hbar{\tau_{\rm intra}(E)}=\frac{|E|}{\frac \zeta 2+\sqrt{1+\frac{E^4}{ E_{i}^4}}-1}, \label{gam-all-E-new}
\ee
where $\zeta$ is a number of order unity. Substituting the model expression (\ref{gam-all-E-new}) into the equilibrium ($\mu_e=\mu_h=\mu_0$) static ($\hbar \omega=0$) zero-temperature ($T=0$, $\mu_0=E_F$) intra-band conductivity (\ref{cond-intra-e}) or (\ref{cond-intra-h}) we obtain
\be
\frac{\sigma^{(1)}_{\rm intra}(0,E_F,0)}{(e^2/h) }
= \zeta+2\left(\sqrt{1+\frac{E_F^4}{ E_{i}^4}}-1\right)=\zeta+2\left(\sqrt{1+\left(\beta  \frac{n_s}{N_i}\right)^2}-1\right),
\label{sigma-fit}
\ee
where $n_s=k_F^2/\pi=E_F^2/\pi(\hbar v_F)^2$ is the charge carrier density at $T=0$ and 
\be 
\beta =\left(\alpha r_s\right)^{-2}\approx 9.61
\ee
is another numerical factor. As seen from (\ref{sigma-fit}) the factor $\zeta$ determines the minimal conductivity of graphene at the Dirac point; in typical experiments $\zeta\simeq 4$. At large densities, $\beta n_s\gg N_i$, the conductivity in (\ref{sigma-fit}) is proportional to $n_s$, $\sigma^{(1)}_{\rm intra}(0,E_F,0)=e\mu n_s$, which gives the relation between the low-temperature mobility $\mu$ and the density of impurities,
\be 
\mu=\frac\beta\pi\frac e{\hbar N_i}.\label{mobility}
\ee
Figure \ref{fig:param} illustrates the relations (\ref{coul-en-1}) and (\ref{mobility}) between the impurity density $N_i$, the energy $E_i$ and the mobility $\mu$.

\begin{figure}[ht]
\includegraphics[width=0.48\textwidth]{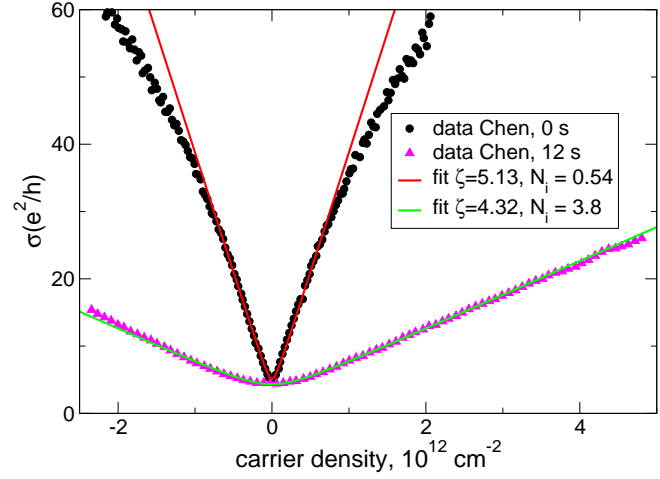}
\caption{\label{fig:Chen08}Experimental data from Ref. \cite{Chen08} for two different potassium atoms densities, corresponding to 0 s and 12 s doping time, see Fig. 2 in Ref. \cite{Chen08}. Symbols -- experimental data, solid curves -- fitting curves (\ref{sigma-fit}) for parameters $\zeta$ and $N_i$ indicated on the plot. The density of charge carriers $n_s$ and of impurities $N_i$ are in units $10^{12}$ cm$^{-2}$, the conductivity is in units $e^2/h$. }
\end{figure}

Varying two adjustable parameters $\zeta$ and $E_i$ one can now fit the expression (\ref{sigma-fit}) to experimental data on the density (or gate-voltage) dependence of the intra-band \textit{static} conductivity and then use thus found parameters $\zeta$ and $E_i$ for calculations of the \textit{high-frequency} linear and nonlinear response. In Figure \ref{fig:Chen08} we illustrate this procedure by fitting (\ref{sigma-fit}) to some of the experimental data from Ref. \cite{Chen08}. In that paper the authors measured the graphene conductivity as a function of gate voltage in samples intentionally doped by potassium atoms. The black and magenta symbols in Figure \ref{fig:Chen08} show the data from Figure 2 of Ref. \cite{Chen08}, for pristine graphene (denoted as 0 s) and for the same sample after the doping during 12 s. The data from Ref. \cite{Chen08} are replotted as a function of charge carrier density
\be 
n_s=c_g(V_g-V_{g0})/e,
\ee
where the gate capacitance per unit area $c_g=1.15\times 10^{-4}$ F/m$^2$ is taken from Ref. \cite{Chen08}. 
One sees that the curves (\ref{sigma-fit}) excellently reproduce experimental data at reasonable values of the fitting parameters $\zeta$ and $N_i$. The found values of $N_i$ are about $N_i\approx 0.54\times 10^{12}$ cm$^{-2}$ for pristine graphene (black symbols) and $N_i\approx 3.8\times 10^{12}$ cm$^{-2}$ for the doping time of 12 s (magenta symbols). The maximum impurity density (for 18 s doping time) was estimated in Ref. \cite{Chen08} as $(1.4-1.8)\times 10^{-3}$ potassium per carbon, or $N_i^{\rm max}\simeq (5.3 - 6.9)\times 10^{12}$ cm$^{-2}$, which agrees very well with the numbers obtained from our fit.  

In the main text we use the energy $E_i$ as a fitting parameter, instead of $N_i$. If $N_i$ varies in the range $\sim (0.5-5) \times 10^{12}$ cm$^{-2}$, the energy $E_i$ lies in the interval from $\sim 26.5$ to $\sim 83.9$ meV and the mobility in the interval from $9300$ to $930$ cm$^2$/Vs. 

%

\end{document}